\documentclass[letterpaper]{article}

\pdfoutput=1
\usepackage{dcolumn}
\usepackage{bm}

\usepackage{graphicx}
\usepackage{amssymb,amsmath}
\usepackage{multirow}
\usepackage{multicol}
\usepackage{widetext}
\usepackage{nonfloat}
\usepackage{cite,color,url}
\usepackage[colorlinks=true
,urlcolor=blue
,anchorcolor=blue
,citecolor=blue
,filecolor=blue
,linkcolor=blue
,menucolor=blue
,linktocpage=true
,pdfproducer=medialab
,pdfa=true
]{hyperref}

\usepackage{epsfig,psfrag,rotating,soul}
\usepackage{rotfloat}
\usepackage{bbold} 

\evensidemargin \oddsidemargin
\allowdisplaybreaks

\psfrag{lk}{$l_k$}
\psfrag{lm}{$l_m$}
\psfrag{blk}{$\bar {l}_k$}
\psfrag{blm}{$\bar {l}_m$}

\def\thefootnote{\fnsymbol{footnote}}

\let\OLDthebibliography\thebibliography
\renewcommand\thebibliography[1]{
  \OLDthebibliography{#1}
  \setlength{\parskip}{0pt}
  \setlength{\itemsep}{0pt plus 0.3ex}
}

\begin{document}

\begin{flushright}
IFT-UAM/CSIC-15-085\\
FTUAM-15-25\\
SU-HET-07-2015\\
IPPP/15/67\\
DCPT/15/134
\end{flushright}

\vspace{0.5cm}

\begin{center}

\begin{Large}
\textbf{\textsc{Enhancement of the lepton flavor violating Higgs boson decay rates from SUSY loops in the inverse seesaw model}}
\end{Large}

\vspace{1cm}

{\sc
E. Arganda$^{1}$%
\footnote{\tt \href{mailto:ernesto.arganda@unizar.es}{ernesto.arganda@unizar.es}}%
, M.J. Herrero$^{2}$%
\footnote{\tt \href{mailto:maria.herrero@uam.es}{maria.herrero@uam.es}}%
, X. Marcano$^{2}$%
\footnote{\tt \href{mailto:xabier.marcano@uam.es}{xabier.marcano@uam.es}}%
, C. Weiland$^{2,3,4}$%
\footnote{{\tt \href{mailto:cedric.weiland@durham.ac.uk}{cedric.weiland@durham.ac.uk.}}}%
}

\vspace*{.7cm}

{\sl
$^1$Departamento de F\'{\i}sica Te\'orica, Facultad de Ciencias,\\
Universidad de Zaragoza, E-50009 Zaragoza, Spain

\vspace*{0.1cm}

$^2$Departamento de F\'{\i}sica Te\'orica and Instituto de F\'{\i}sica Te\'orica, IFT-UAM/CSIC,\\
Universidad Aut\'onoma de Madrid, Cantoblanco, 28049 Madrid, Spain

\vspace*{0.1cm}

$^3$Graduate School of Science and Engineering,\\
Shimane University, Matsue, 690-8504 Japan 

\vspace*{0.1cm}

$^4$Institute for Particle Physics Phenomenology,\\
Department of Physics, Durham University, Durham DH1 3LE, United Kingdom

}

\end{center}

\vspace*{0.1cm}

\begin{abstract}
\noindent
In this article we study the full one-loop SUSY contributions to the lepton flavor violating Higgs decay $h \to \tau \bar \mu $, within the context of the supersymmetric inverse sesaw model. We
assume that both the right-handed neutrino masses, $M_R$, and their supersymmetric partner masses, $m_{\tilde \nu_R}$, are not far from the interesting ${\cal O}({\rm TeV})$ energy
scale, and we work with scenarios with large neutrino Yukawa couplings that transmit large lepton flavor violating effects. By exploring  the behavior with the most relevant
parameters, mainly $M_R$, $m_{\tilde \nu_R}$ and the trilinear sneutrino coupling $A_\nu$, we will look for regions of the parameter space where the enhancement
of  $\mathrm{BR}(h\rightarrow \tau \bar \mu)$ is large enough to reach values at the percent level, which could explain the excess recently reported by CMS and ATLAS at the CERN Large
Hadron Collider.   
\end{abstract}

\vspace*{0.5cm}

\def\thefootnote{\arabic{footnote}}
\setcounter{footnote}{0}

\begin{multicols}{2}

\section{Introduction}
\label{intro}

The discovery in 2012 of a new scalar particle at the LHC~\cite{Aad:2012tfa,Chatrchyan:2012ufa}, whose mass has been set to $m_h =$ 125.09 $\pm$ 0.21 (stat.) $\pm$ 0.11 (syst.) GeV~\cite{Aad:2015zhl}, lays on the table the challenging issue of whether it is actually the Higgs boson
from the standard model of particle physics (SM) or there is new physics beyond the SM (BSM).

In this article, we focus on one of the new physics aspects of the discovered boson---the possibility of lepton flavor violating Higgs decays (LFVHD). In fact, very recently, 
the first direct search of the particular decay $h \to \mu \tau$ (from now on, we will refer to both $h \to \mu \bar \tau$ and $h \to \tau \bar \mu$ decays in this shortened way) has been performed by the
CMS Collaboration~\cite{Khachatryan:2015kon}, and an upper limit on the branching ratio of BR($h \to \mu \tau$) $< 1.51 \times 10^{-2}$ at 95\% C.L. has been
set. Additionally, CMS has also observed a slight excess with a significance of 2.4 standard deviations at $m_h =$ 125 GeV, whose best-fit branching
ratio, if interpreted as a signal, is BR($h \to \mu \tau$) $= (8.4_{-3.7}^{+3.9}) \times 10^{-3}$. The ATLAS Collaboration has just released their results for the
same $h \to \mu \tau$ decay~\cite{Aad:2015gha} as well, focusing on hadronically decaying $\tau$ leptons. ATLAS has reported an upper limit of BR($h \to \mu \tau$) $< 1.85 \times 10^{-2}$ at 95\% C.L. in agreement with the previous CMS result. 
Intriguingly, a small excess appears in one of the signal regions considered, even though it is not statistically significant.
One way or another, the searches for lepton flavor violation (LFV) in the Higgs
sector have entered into the percent level. The statistical significance is not enough to reach a strong conclusion yet, but any evidence of LFV would unquestionably mean a
clear BSM signal due to the huge suppression of LFV in the SM because of
the absence of flavor-changing neutral currents.

In particular, the investigation of LFVHD is at present a very active field which is being studied in different models.
LFVHD were considered for the first time
in the context of
the SM enlarged with three heavy Majorana neutrinos in~\cite{Pilaftsis:1992st} and later in the context of the type I seesaw model in~\cite{Arganda:2004bz}, predicting tiny rates
due to the strong suppression from the large heavy right-handed neutrino masses. By contrast, in the context of the inverse seesaw model (ISS)~\cite{ISSrefs} with right-handed
neutrino masses at the ${\cal O}({\rm TeV})$ energy scale, much larger LFVHD rates, up to $10^{-5}$, can be obtained~\cite{Arganda:2014dta}. In addition, LFVHD have been also
analyzed with special attention in the literature within the framework of supersymmetric (SUSY) models~\cite{Arganda:2004bz,LFVHDsusy,Arana-Catania:2013xma}, finding branching
ratios slightly larger
than in the ISS case, up to $10^{-4}$. 

Here we will study the LFVHD within the context of the SUSY version of the ISS, which we refer to here as the SUSY-ISS model. In particular, we will present our estimate of the contribution
to the BR($h \to \tau \bar \mu$) from all the SUSY loops containing sneutrinos and sleptons which are typically different in the SUSY-ISS with respect to other SUSY models, due to
the important effects induced by the right-handed neutrinos and their SUSY partners with masses at ${\cal O}({\rm TeV})$. The potential increase of the LFVHD rates due to some
of the new SUSY loops within the SUSY-ISS model was first pointed out and estimated in~\cite{Abada:2011hm}. Other important enhancement due to SUSY loops have also been found
in~\cite{Abada:2014kba} for LFV lepton decay rates and other observables. Some phenomenological implications at the LHC of SUSY-ISS scenarios with large LFVHD rates within
the same context as this work have been recently studied in \cite{Arganda:2015ija}.

In addition to performing a complete one-loop computation of the SUSY loops within the SUSY-ISS model, one of our main goals here is to analyze in detail if the enhancement due to
the sneutrinos and sleptons loops can be sufficiently large as to explain the LFVHD effect seen by CMS and ATLAS. Indeed, we will localize in this work some regions of the SUSY-ISS parameter
space where this is possible. In section~\ref{th-framework} we describe the SUSY-ISS model and introduce the parametrization we use to reproduce low-energy neutrino data. In
section~\ref{Sec:Analytical} we present the analytical results of our one loop calculation while we discuss our numerical predictions in section~\ref{results}.
 
\section{The SUSY-ISS Model}
\label{th-framework}

In this section, we briefly summarize the most relevant aspects for the present computation of the SUSY-ISS model, which is a well-known extension of the MSSM that can reproduce the observed
neutrino masses and mixing. The MSSM superfield content is supplemented by three pairs of gauge singlet chiral superfields $\widehat N_i$ and $\widehat X_i$ with opposite lepton
numbers ($i=1,2,3$). The SUSY-ISS model is defined by the following superpotential:
\begin{equation}
 W=W_\mathrm{MSSM} + \varepsilon_{ab} \widehat N Y_\nu \widehat H^b_2 \widehat L^a + \widehat N \widetilde M_R \widehat X + \frac{1}{2} \widehat X \widetilde \mu_X \widehat X\,,
\end{equation}
with $\varepsilon_{12}=1$ and
\begin{align}
 W_\mathrm{MSSM}=&\varepsilon_{ab} \left[ \widehat E Y_e \widehat H^a_1 \widehat L^b + \widehat D Y_D \widehat H^a_1 \widehat Q^b + \widehat U Y_U \widehat H^b_2 \widehat Q^a\right. \nonumber \\
	  &\left.- \mu \widehat H^a_1 \widehat H^b_2 \right]\,.
\end{align}
The generation indices have been suppressed and should be understood in a tensor notation as
$\widehat N Y_\nu \widehat H^b_2 \widehat L^a=\widehat N_i (Y_\nu)_{ij} \widehat H^b_2 \widehat L^a_j$. In particular, all chiral superfields
are left-handed, meaning that for $\widehat D\,, \widehat U\,, \widehat E\,, \widehat N\,, \widehat X$ the spin $0$ and spin $\frac {1}{2}$ components are, for example in the case of
$\widehat E$, $[(\widetilde{e_R})^*\,, (e_R)^c]$. $\widehat H_1$ and $\widehat H_2$ are, respectively, the down-type and up-type Higgs bosons, defined as
\begin{equation}
 \widehat H_1 = \binom{\hat h^0_1}{\hat h^-_1}\,,	\quad \widehat H_2 = \binom{\hat h^+_2}{\hat h^0_2}\,.
\end{equation}

Then the soft SUSY breaking Lagrangian is given by
\begin{align}
 -\mathcal{L}_\mathrm{soft}=&-\mathcal{L}_\mathrm{soft}^\mathrm{MSSM} + \widetilde \nu_R^T\, m_{\tilde \nu_R}^2\, \widetilde \nu_R^* + \widetilde X^T m_{\tilde X}^2 \widetilde X^*\\ \nonumber
		      & + \widetilde \nu_R^\dagger (A_{\nu} Y_\nu) \widetilde \nu_L h_2^0 - \widetilde \nu_R^\dagger (A_{\nu} Y_\nu) \widetilde e_L h_2^+ + \mathrm{h.c.}\\ \nonumber
		      & + \widetilde X^\dagger (B_{X} \widetilde \mu_X) \widetilde X^* + \widetilde \nu_R^\dagger (B_{R} \widetilde M_R) \widetilde X^* + \mathrm{h.c.}\,,
\end{align}
with
\begin{align}
 -\mathcal{L}_\mathrm{soft}^\mathrm{MSSM}=&~ \widetilde e_R^T\, m_{\tilde e}^2\, \widetilde e_R^* +  \widetilde d_R^T\, m_{\tilde d}^2\, \widetilde d_R^* + \widetilde u_R^T\, m_{\tilde u}^2\, \widetilde u_R^* \\
		      &+ m_{H_1}^2 |H_1|^2 + m_{H_2}^2 |H_2|^2 \nonumber \\
		      & + \delta_{ab} (\widetilde Q^a)^\dagger m_{\tilde Q}^2 \widetilde Q^b + \delta_{ab} (\widetilde L^a)^\dagger m_{\tilde L}^2 \widetilde L^b \nonumber \\
		      & + \frac{1}{2} \left( M_1 \bar \lambda_b \lambda_b + M_2 \bar \lambda^\alpha_W \lambda^\alpha_W \right. \nonumber \\
		      &  \quad + \left. M_3 \bar \lambda^\alpha_g \lambda^\alpha_g + \mathrm{h.c.}\right) \nonumber \\
		      & + \varepsilon_{ab} \left[ (\widetilde u_R^\dagger (A_u Y_u) \widetilde Q^a H^b_2\right. \nonumber \\
		      & \quad \left.+ \widetilde d_R^\dagger (A_d Y_d) \widetilde Q^b H^a_1\right. \nonumber \\
		      & \quad \left.+ \widetilde e_R^\dagger (A_e Y_e) \widetilde L^b H^a_1 + B \mu H_2^a H_1^b + \mathrm{h.c.} \right]. \nonumber
\end{align}
During this study we will take all soft SUSY breaking masses to be flavor diagonal, making sure that the only sources of flavor violation are
in the neutrino Yukawa coupling $Y_\nu$, the lepton number conserving mass term $\widetilde{M_R}$ and the lepton number violating mass term $\widetilde \mu_X$. The only
exception will be $m_{\tilde L}^2$ which receives RGE-induced corrections coming from $Y_\nu$ that, for phenomenological purposes, are given by~\cite{Hisano:1995cp}
\begin{equation} \label{SleptonMixing}
 (\Delta m_{\tilde L}^2)_{ij} = -\frac{1}{8\pi^2} (3 M_0^2 + A_0^2) (Y_\nu^\dagger \log  \frac{M}{M_R} Y_\nu)_{ij}\,,
\end{equation}
where we take $M=10^{18}\,\mathrm{GeV}$ for the rest of this work.

After electroweak symmetry breaking, the neutrino mass matrix in the basis $((\nu_L)^c\,,\;\nu_R\,,\;X)^T$ is given by
\begin{equation}
 M_{\mathrm{ISS}} = \left(\begin{array}{c c c} 0 & m_D & 0 \\ m_D^T & 0 & M_R \\ 0 & M_R^T & \mu_X \end{array}\right)\,,
\end{equation}
where we have defined $m_D = Y_\nu^\dagger v_2$, with $v_2=<h^0_2>$, $M_R = \widetilde M_R^*$ and $\mu_X = \widetilde \mu_X^*$ in order to agree with the definitions used in our 
previous article on
LFV Higgs decays~\cite{Arganda:2014dta}. In the limit $\mu_X \ll m_D \ll M_R$, it is possible to diagonalize by blocks this matrix~\cite{GonzalezGarcia:1988rw}, leading to
the $3 \times 3$ light neutrino mass matrix
\begin{equation}
M_{\mathrm{light}} \simeq m_D {M_R^T}^{-1} \mu_X M_R^{-1} m_D^T\,,
\end{equation}
which, in turn, is diagonalized by the PMNS matrix $U_{\rm PMNS}$\cite{PMNS}:
\begin{equation} \label{mnulight}
 U_{\rm PMNS}^T M_{\mathrm{light}} U_{\rm PMNS} =m_\nu\,,
\end{equation}
where $m_\nu=\mathrm{diag}(m_{\nu_1}\,, m_{\nu_2}\,, m_{\nu_3})$ is the diagonal matrix that contains the masses of the three lightest neutrinos. Low-energy neutrino data can be
reproduced by using the following parametrization introduced in \cite{Arganda:2014dta}: 
\begin{equation}
\mu_X=M_R^T ~m_D^{-1}~ U_{\rm PMNS}^* m_\nu U_{\rm PMNS}^\dagger~ {m_D^T}^{-1} M_R\,.
\end{equation}

In particular, this parametrization allows us to use the neutrino Yukawa couplings $Y_\nu$ as input parameters. As we showed in~\cite{Arganda:2014dta}, the following three $Y_\nu$
textures
\begin{align}
Y_{\tau \mu}^{(1)}=f\left(\begin{array}{ccc}
0&1&-1\\0.9&1&1\\1&1&1
\end{array}\right)\,, \\
Y_{\tau \mu}^{(2)}=f\left(\begin{array}{ccc}
0&1&1\\1&1&-1\\-1&1&-1
\end{array}\right)\,, \\
Y_{\tau \mu}^{(3)}=f\left(\begin{array}{ccc}
0&-1&1\\-1&1&1\\0.8&0.5&0.5
\end{array}\right)\,,
\end{align}
where $f$ is a scaling factor, can lead to large $\tau-\mu$ flavor transition rates while suppressing $\mu-e$ and $\tau-e$ flavor transition rates. We found that in the
nonsupersymmetric ISS model,
these could lead to large branching ratios for LFV Higgs decays, up to $10^{-5}$, while still agreeing with other experimental constraints. As mentioned in the introduction, previous
studies have demonstrated that supersymmetric contributions usually enhance the LFV rates. In particular, in the present SUSY-ISS model, since we consider a seesaw
scale $M_R$ not far from the electroweak scale, this low value will enhance the flavor slepton
mixing due to the RGE-induced radiative effects by the large neutrino Yukawa couplings, and this mixing will in turn generate via the slepton loops an enhancement in the LFVHD
rates. On the other hand, new relevant couplings appear, like $A_{\nu}$, which for right-handed sneutrinos with $\mathcal{O}(1\,\mathrm{TeV})$ masses may lead to new loop
contributions to LFVHD
that could even dominate~\cite{Abada:2011hm}. In light of the recent CMS and ATLAS searches~\cite{Khachatryan:2015kon,Aad:2015gha} for $h \to \mu \tau$, this calls for a new and complete evaluation 
of the SUSY contributions to this observable in the SUSY-ISS model.

\section{Analytical results}
\label{Sec:Analytical}

In this work, we perform a full one-loop diagrammatic computation of all relevant supersymmetric loops within the SUSY-ISS model for $\mathrm{BR}(H_x\rightarrow \ell_k \bar \ell_m)$,
where $H_x$ here and from now on refers to the three neutral MSSM Higgs bosons, $H_x=(h,H,A)$. 
This is in contrast to the previous estimate
in~\cite{Abada:2011hm} where an effective Lagrangian description of the Higgs mediated contributions to LFV processes was used, which was appropriate to capture the relevant
contributions at large $\tan\beta$, and where the mass insertion approach was
used to incorporate easily 
the flavor slepton mixing $(\Delta m_{\tilde L}^2)_{ij}$, working in the electroweak basis. However, an expansion up to the first order in the mass insertion approximation
may not be appropriate for the type of scenarios studied here,
due to the large flavor-nondiagonal matrix entries considered in this work. On the other hand, we are interested also in small and moderate $\tan\beta$ values, not just in the
large $\tan\beta$ regime, and we also wish to explore more generic soft masses for the SUSY particles and scan over the relevant neutrino/sneutrino parameters,
mainly $M_R$, $A_\nu$ and $m_{{\tilde \nu}_R}$, not focusing only on scenarios with universal or partially universal soft parameters nor fixing the relevant parameters to one
value as in~\cite{Abada:2011hm}. Thus, our calculation is performed instead in the mass basis for all the SUSY particles involved in the loops, i.e,
the charged sleptons, sneutrinos, charginos, and neutralinos.

Before moving to the calculation, let us introduce the relevant interaction terms from the Lagrangian for the study of the LFV Higgs decays.
Following the notation in~\cite{Arganda:2004bz}, these terms are given in the mass basis by 
\begin{align}
\mathcal{L}_{\tilde \chi_j^- \ell \tilde \nu_{\alpha} } &= 
-g\, \bar{\ell} \left[ A_{L \alpha j}^{(\ell)} P_L + 
A_{R \alpha j}^{(\ell)} P_R \right] \tilde \chi_j^- \tilde \nu_{\alpha} + h.c.\,, \nonumber \\
\mathcal{L}_{\tilde \chi_a^0 \ell \tilde \ell_{\alpha} } &= 
-g\bar{\ell} \left[ B_{L \alpha a}^{(\ell)} P_L + 
B_{R \alpha a}^{(\ell)} P_R \right] \tilde \chi_a^0 \tilde \ell_{\alpha} + h.c.\,, \nonumber \\ 
\mathcal{L}_{H_x \tilde s_{\alpha} \tilde s_{\beta}}&=
-\imath H_x \left[ g_{H_x\tilde \nu_{\alpha} \tilde \nu_{\beta}} \tilde \nu_{\alpha}^* 
\tilde \nu_{\beta}
+ g_{H_x\tilde \ell_{\alpha} \tilde \ell_{\beta}} \tilde \ell_{\alpha}^* \tilde \ell_{\beta}
\right]\,,\nonumber\\
\mathcal{L}_{H_x \tilde \chi_i^- \tilde \chi_j^-}&=
- g H_x\bar{\tilde{\chi}}_i^- 
\left[ W_{Lij}^{(x)}P_L+ W_{Rij}^{(x)}P_R \right] \tilde{\chi}_j^-\,,\nonumber\\
\mathcal{L}_{H_x \tilde \chi_a^0 \tilde \chi_b^0}&=
- \frac{g}{2} H_x\bar{\tilde{ \chi}}_a^0 
\left[ D_{Lab}^{(x)}P_L+ D_{Rab}^{(x)}P_R \right] \tilde {\chi}_b^0\,, \nonumber \\
\mathcal{L}_{H_x\ell\ell} &= 
-g H_x\bar{\ell} \left[ S_{L,\ell}^{(x)} P_L + S_{R,\ell}^{(x)} P_R \right]  \ell \,,
\end{align}
where the coupling factors have been expressed in terms of the SUSY-ISS model parameters and are collected in Appendix~\ref{App_Couplings}.

\begin{figure*}[htb]
\begin{center}
\includegraphics[width=\textwidth]{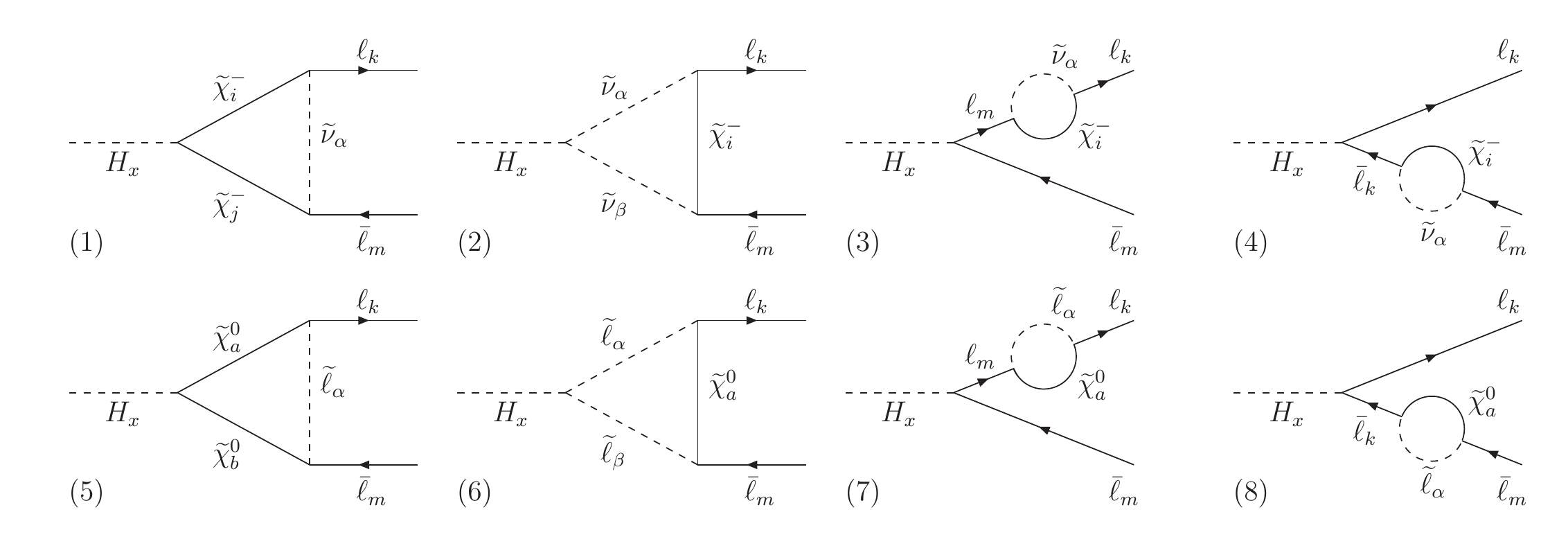}
\caption{One-loop supersymmetric diagrams contributing to the process $H_x \rightarrow \ell_k \bar \ell_m$.}
\label{SUSY-diag}
\end{center}
\end{figure*}

We take into account the full set of 1-loop SUSY diagrams shown in figure~\ref{SUSY-diag}.
It is interesting to notice that, since we work in the mass basis, the set
of diagrams contributing to LFV Higgs decays (four diagrams with charginos and sneutrinos in the loops, and four more with neutralinos and charged sleptons) is the same as in
the SUSY type I seesaw model which was considered in~\cite{Arganda:2004bz}. We keep their definition of the form factors
\begin{equation}
\imath F_x = -\imath g \bar{u}_{\ell_k} (-p_2) (F_{L,x} P_L + F_{R,x} P_R) v_{\ell_m}(p_3)\,,
\end{equation}
where $F_x$ is the decay amplitude for $H_x \rightarrow \ell_k \bar \ell_m$ with again $H_x=(h,H,A)$ and $p_1=p_3-p_2$ is the ingoing Higgs boson momentum.
The contributions of the SUSY diagrams are summed in $F_{L,x}$ and $F_{R,x}$ according to
\begin{equation}
F_{L,x} = \sum_{i=1}^{8} F_{L,x}^{(i)},\,\,  F_{R,x} = \sum_{i=1}^{8} F_{R,x}^{(i)}\,.
\end{equation}
Their analytic expressions are taken
from~\cite{Arganda:2004bz} and reproduced in Appendix~\ref{App_FF} for completeness, including the proper modifications to adapt them to the SUSY-ISS model.
We have checked analytically the cancellation of divergences appearing in the loop contributions in both form factors $F_{L,x}$ and $F_{R,x}$ in eq.(16),giving, as expected, a finite result without the need of the renormalization procedure. Notice that this cancellation is not trivial and is, therefore, a good test of our results of the form factors in Appendix~\ref{App_FF}. 
The parametrization of the LFVHD widths in terms of form factors remains unchanged and is given by
\begin{align}
&\Gamma (H_x \rightarrow {\ell_k} \bar{\ell}_m) = \frac{g^2}{16 \pi m_{H_x}} \left(- 4 m_{\ell_k} m_{\ell_m} Re(F_{L,x} F_{R,x}^*)\right. \nonumber \\ 
&\left. + (m_{H_x}^2-m_{\ell_k}^2-m_{\ell_m}^2)(|F_{L,x}|^2+|F_{R,x}|^2) \right) \\ 
&\times \sqrt{\left(1-\left(\frac{m_{\ell_k}+m_{\ell_m}}{m_{H_x}}\right)^2\right)
\left(1-\left(\frac{m_{\ell_k}-m_{\ell_m}}{m_{H_x}}\right)^2\right)}\,, \nonumber
\end{align}
where $g$ is the $\mathrm{SU}(2)_L$ coupling constant, $m_{H_x}$ is the mass of the Higgs boson while $m_{\ell_k}$ and $m_{\ell_m}$ are the masses of final state leptons.

\section{Numerical Results}
\label{results}

In this section we show the numerical results of the LFV decay rates of the lightest neutral Higgs boson, BR($h \to \tau \bar\mu$), as a function of the most relevant parameters
of the SUSY-ISS model for the full SUSY contribution to LFVHD, namely,
$M_R$, $A_\nu$ and $m_{{\tilde \nu}_R}$. 
It should be noted that, in the absence of CP violation, as in our case, BR($h\to\tau\bar\mu$)=BR($h\to\mu\bar\tau$) and, therefore, in comparing with data the two rates should be added.

We have imposed various experimental constraints, choosing as example two benchmark points
leading to a Higgs
boson mass within 1$\sigma$ of the central value of the latest CMS and ATLAS combination, and with supersymmetric spectrum allowed by ATLAS and CMS
searches. Indeed, we concentrate for this work on the slepton sector, since the squark sector is irrelevant for the LFVHD. The squark parameters are only relevant for the Higgs
mass prediction, thus one can always adjust
the squark masses and the trilinear couplings $A_t$ and $A_b$ in order to ensure a correct Higgs boson mass. We also restrict ourselves to the case of $M_R>m_h$, avoiding
constraints from the invisible Higgs decay widths, and consider only real $U_\mathrm{PMNS}$ and
mass matrices, making constraints from lepton electric dipole moments irrelevant. Finally, we also take into account the LFV radiative decays whose current upper limits
at the $90\%$ C.L. are
\begin{align}
{\rm BR}(\mu\to e\gamma)&\leq 5.7\times 10^{-13}\text{\cite{Adam:2013mnn}}\label{MUEGmax}\,,\\
{\rm BR}(\tau\to e\gamma)&\leq 3.3\times 10^{-8}~\text{\cite{Aubert:2009ag}}\label{TAUEGmax}\,,\\
{\rm BR}(\tau\to \mu\gamma)&\leq 4.4\times 10^{-8}~\text{\cite{Aubert:2009ag}}\label{TAUMUGmax}\,.
\end{align}
Points excluded by LFV radiative decays will be denoted by a cross, while a triangle will represent the ones allowed. We present here the predictions
of BR($h \to \tau \bar \mu$) for the
three neutrino Yukawa textures exposed in the section~\ref{th-framework}, ensuring the practically vanishing LFV in the $\mu-e$ sector, i.e., leading
to BR$(\mu \to e \gamma)\sim 0$ and BR$(h \to e \bar \mu)\sim 0$.  It should be noticed that these textures also suppress substantially the LFV in the $\tau - e$ sector. Therefore,
the most stringent constraint, making use of these textures, is that of the related LFV
radiative decay $\tau \to \mu \gamma$.

\begin{figure*}[t!]
\begin{center}
\begin{tabular}{cc}
\includegraphics[width=0.475\textwidth]{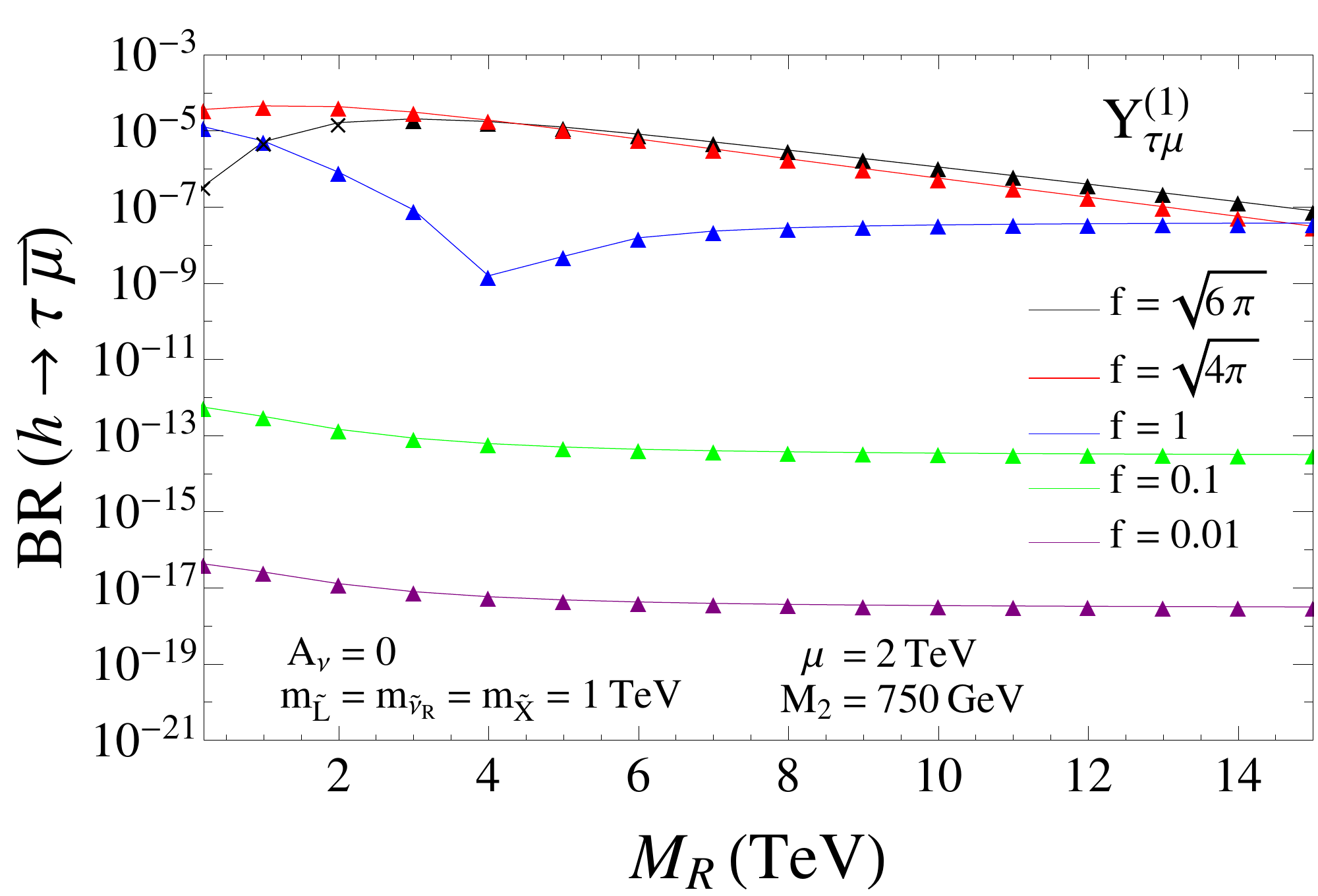}&
\includegraphics[width=0.475\textwidth]{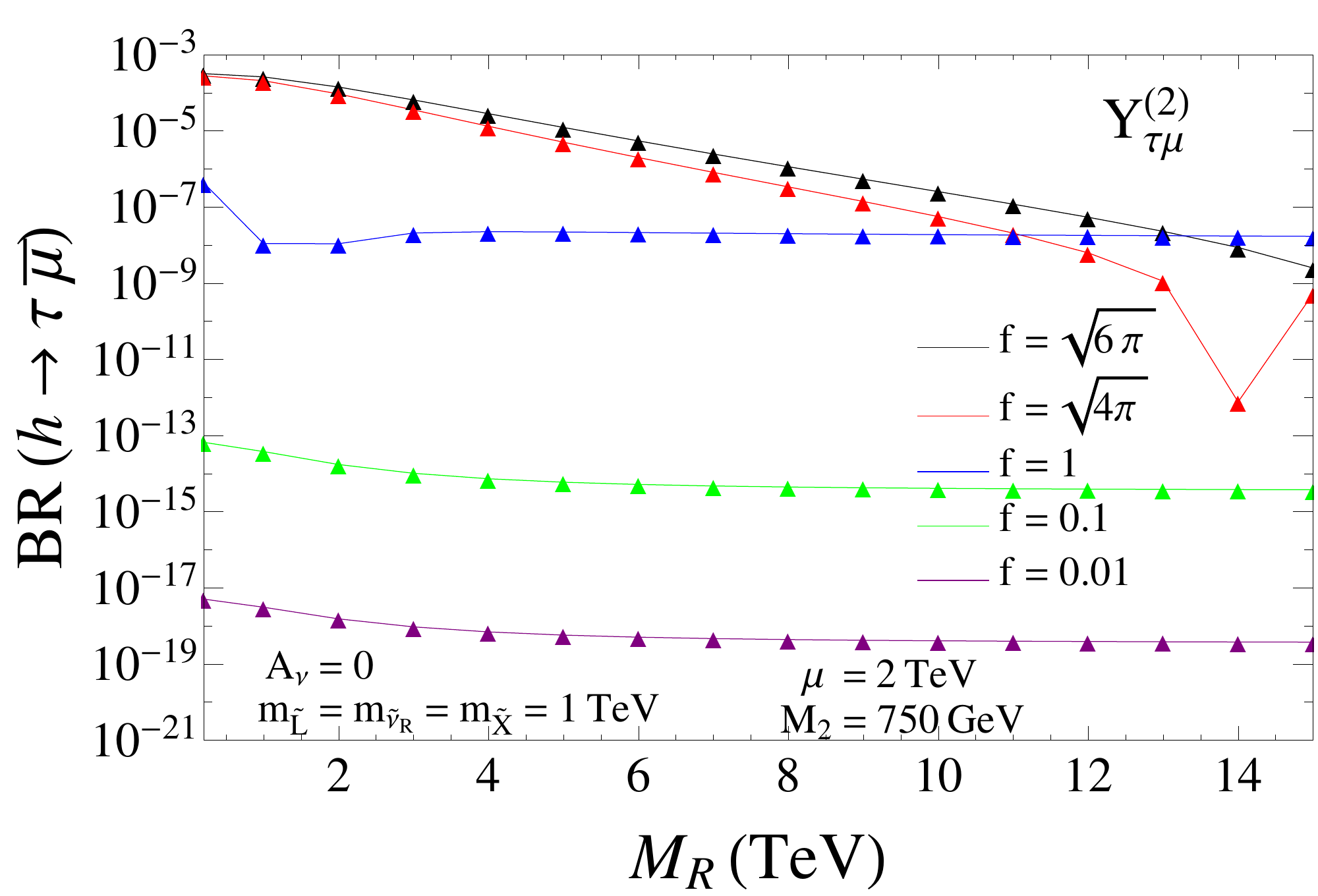}\\
\includegraphics[width=0.475\textwidth]{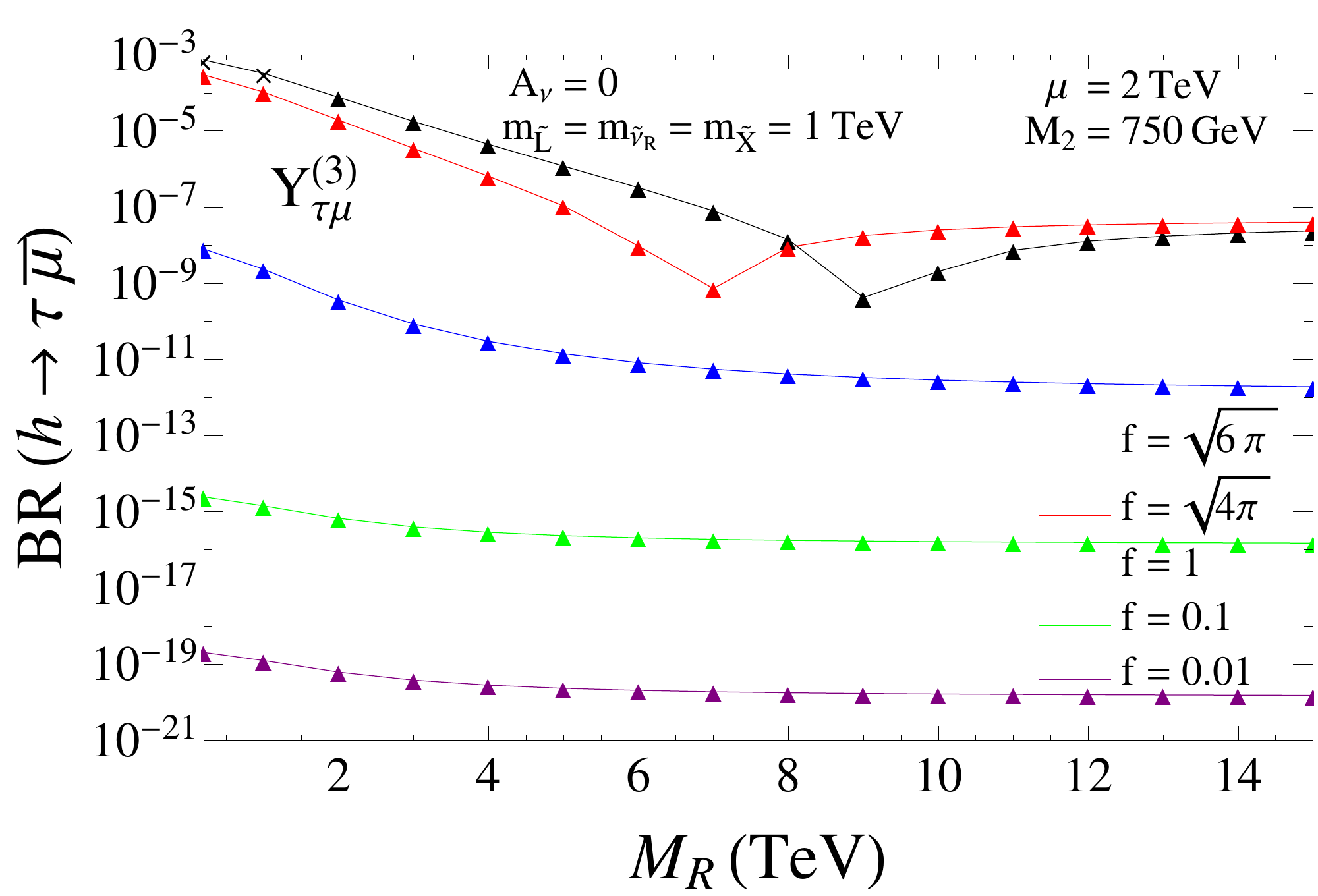}&
\includegraphics[width=0.475\textwidth]{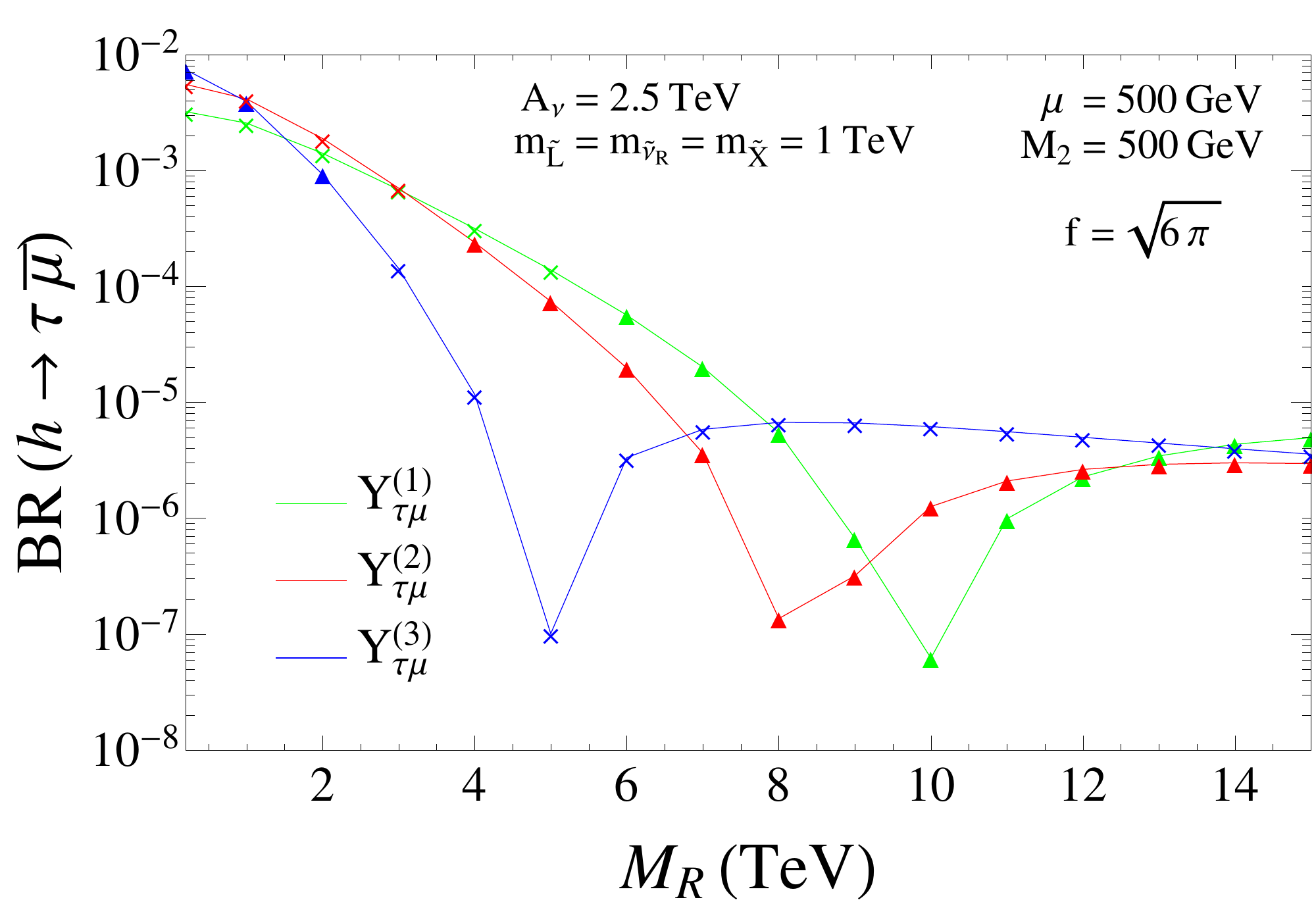}
\end{tabular}
\caption{BR($h \to \tau \bar \mu$) as a function of $M_R$ for $Y_{\tau\mu}^{(1)}$ (upper left panel), $Y_{\tau\mu}^{(2)}$ (upper right panel),
and $Y_{\tau\mu}^{(3)}$ (lower left panel), with $m_{\tilde L} =$ $m_{\tilde e} =$ $m_{\tilde \nu_R} =$ $m_{\tilde X} =$ 1 TeV, $M_2 =$ 750 GeV, $\mu =$ 2 TeV, $A_\nu =$ 0, $\tan\beta =$ 5 and
different values of the scaling factor $f =$ 0.01, 0.1, 1, $\sqrt{4\pi}$, $\sqrt{6\pi}$. On the lower right panel, the behavior of BR($h \to \tau \bar \mu$) as a
function of $M_R$ is shown for the three textures with $m_{\tilde L} =$ $m_{\tilde e} =$ $m_{\tilde \nu_R} =$ $m_{\tilde X} =$ 1 TeV, $M_2 = \mu =$ 500 GeV, $A_\nu =$ 2.5 TeV, $\tan\beta =$ 10
and $f = \sqrt{6\pi}$. On all the panels, $m_A =$ 800 GeV and $M_0=$ 1 TeV. 
We set $A_0=A_e=B_X=B_R=0$ and the GUT inspired relation $M_1=5/3~ M_2\tan^2\theta_W$ in these and all the figures of the paper. Crosses (triangles) represent points in the SUSY-ISS parameter space excluded
(allowed) by the $\tau \to \mu \gamma$ upper limit, BR($\tau \to \mu \gamma$) $< 4.4\times 10^{-8}$~\cite{Aubert:2009ag}.}
\label{LFVHD-MR}
\end{center}
\end{figure*}

In figure~\ref{LFVHD-MR},
we show the behavior of BR($h \to \tau \bar \mu$) as a function of $M_R$ for the three textures presented in the previous
section, $Y_{\tau\mu}^{(1)}$ (upper left panel), $Y_{\tau\mu}^{(2)}$ (upper right panel), and $Y_{\tau\mu}^{(3)}$ (lower left panel), for different values
of the scaling factor $f =$ 0.01, 0.1, 1, $\sqrt{4\pi}$, $\sqrt{6\pi}$. First of all, we clearly see that, as expected, the larger the value of $f$ is, the larger
the LFV rates are. We also observe qualitatively different behaviors of the LFV rates between small ($f < 1$) and large ($f > 1$) neutrino Yukawa couplings. As we have checked, this
difference comes from the different behavior with $M_R$ of the two participating types of loops, the ones with charged sleptons where the LFV is generated exclusively by
the mixing $(\Delta m_{\tilde L}^2)_{ij}$ and the ones with sneutrinos where the LFV is generated by both $(\Delta m_{\tilde L}^2)_{ij}$ and $(Y_\nu)_{ij}$. In the case of
small $f$, charged slepton-neutralino loops dominate and they only depend logarithmically on $M_R$ as can be seen from
eq.~(\ref{SleptonMixing}), leading to the apparent flat behavior. However, we checked that this flat behavior disappears when both $M_R$ and $M_0$ (and as a consequence all slepton
and sneutrino masses) increase simultaneously.
When the scale factor $f$ becomes larger, contributions from sneutrino-chargino loops become sizable and even dominate at
low $M_R$. They decrease with $M_R$, due to the increase in the singlet sneutrino masses, which explains the decrease in BR($h \to \tau \bar \mu$) observed in the
upper plots and on the left-hand side of the bottom plots for large $f>1$. In the latter, the appearance of dips due to negative interferences between the two types of loops marks the transition
between the two regimes, with the main contribution coming from sneutrino-chargino loops at low $M_R$ and from slepton-neutralino loops at large $M_R$.
For the first benchmark point, the largest BR($h \to \tau \bar \mu$), allowed by the $\tau \to \mu \gamma$ upper limit, are obtained for $f =$ $\sqrt{4\pi}$ or
$\sqrt{6\pi}$ and $M_R <$ 2 TeV, with a value of $\mathcal{O}(10^{-4})$ for the three textures, which could be probed in future runs of the LHC. Up to now, the trilinear neutrino
coupling $A_\nu$ had been set to zero, whilst on the lower right panel of figure~\ref{LFVHD-MR} we have chosen $A_\nu =$ 2.5 TeV and show the behavior of 
BR($h \to \tau \bar \mu$) with $M_R$ for the three textures with a scaling factor $f = \sqrt{6\pi}$. This value of $A_\nu$ leads to a suppression of the $\tau\rightarrow\mu\gamma$ decay rates while
simultaneously enhancing BR($h \to \tau \bar \mu$). As a consequence, very large LFVHD branching ratios can be obtained for $Y_{\tau\mu}^{(3)}$ with low $M_R \sim 1$ TeV, allowed
by $\tau \to \mu \gamma$, achieving values up to $7 \times 10^{-3}$. These large rates are very close to the percent level and within the sensitivity of the present experiments.

\begin{figure*}[t!]
\begin{center}
\begin{tabular}{cc}
\includegraphics[width=0.475\textwidth]{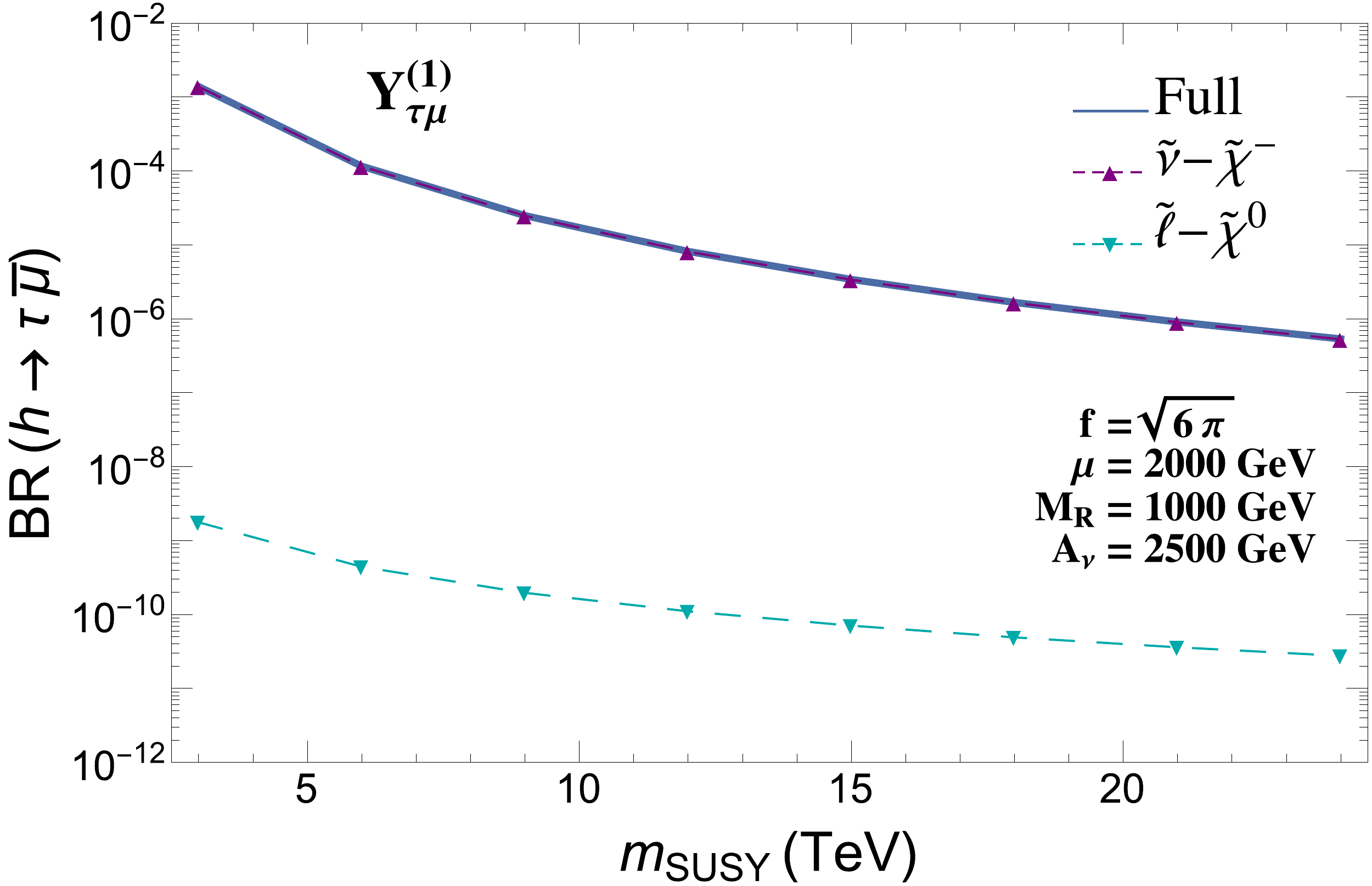}&
\includegraphics[width=0.475\textwidth]{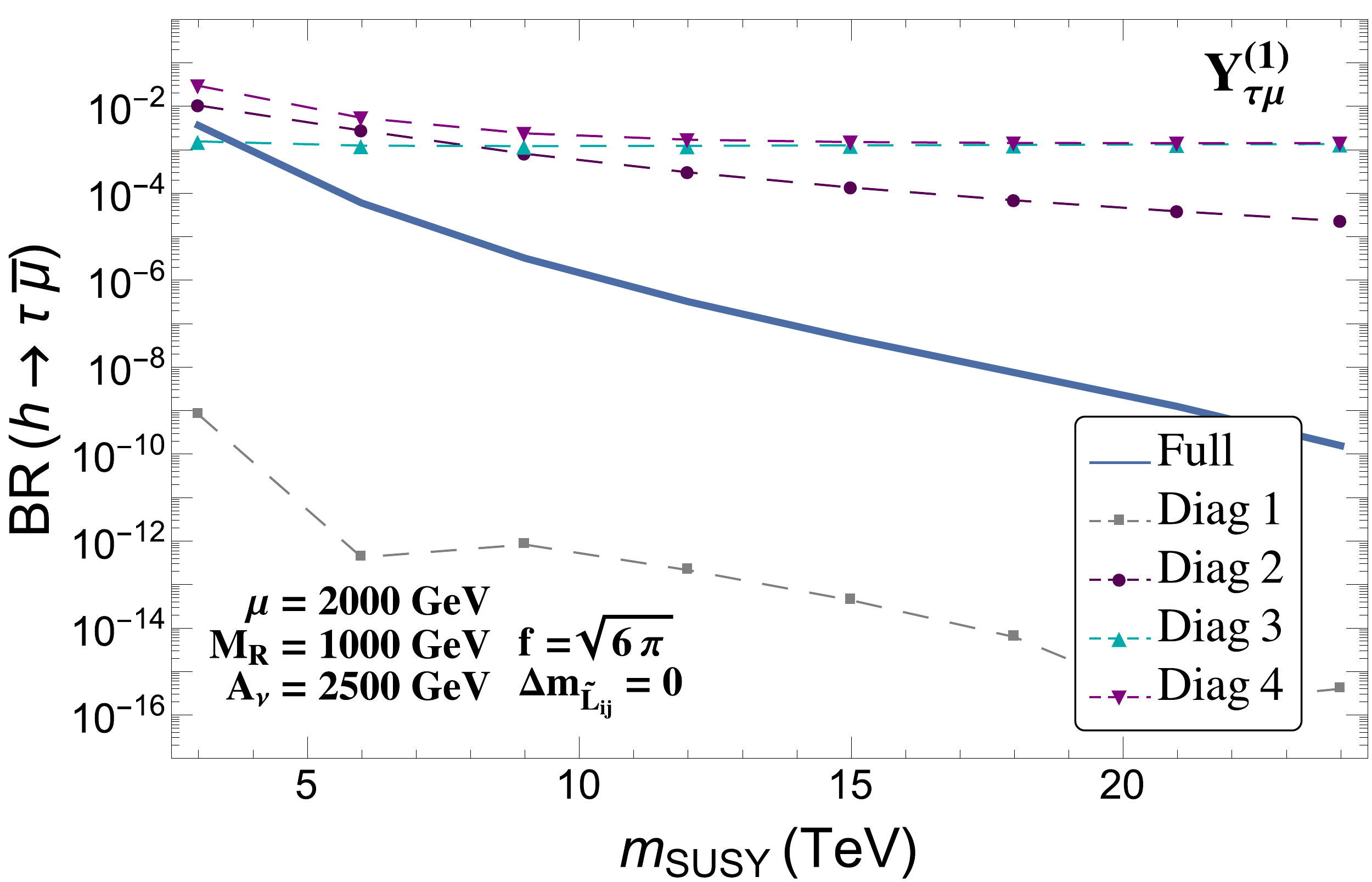}
\end{tabular}
\caption{BR($h\to\tau\bar\mu$) as function of the common SUSY mass parameter $m_{\rm SUSY}$ defined in eq.(\ref{msusy}) for the Yukawa coupling matrix $Y_{\tau\mu}^{(1)}$ with $M_R=1$~TeV, $f=\sqrt{6\pi}$,  $m_A=800$~GeV, $\mu=2$~TeV, $\tan\beta=10$ and $A_\nu=2.5$~TeV.
Left panel: Contributions from chargino-sneutrino loops, denoted by $\widetilde\nu$-$\widetilde\chi^-$, neutralino-slepton loops, denoted by $\widetilde\ell$-$\widetilde\chi^0$, and full results for BR($h\to\tau\bar\mu$). Right panel: Individual contributions from each chargino-sneutrino diagram (1), (2), (3), and (4) in fig.\ref{SUSY-diag}  and full result in the case of $\Delta m_{\tilde L_{ij}}=0$, where the neutralino-slepton contributions vanish.}\label{decoulingplots}
\end{center}
\end{figure*}
 
We next study the behavior of BR($h\to\tau\bar\mu$) as function of the SUSY mass scales in a simplified scenario where all the SUSY masses are equal to a common parameter $m_{\rm SUSY}$, namely,
\begin{equation}\label{msusy}
m_{\rm SUSY}=m_{\tilde L}=m_{\tilde e}=m_{\tilde \nu_R}=m_{\tilde X}=M_0=M_1=M_2.
\end{equation}
Fig.\ref{decoulingplots} left shows the expected decoupling behavior where BR($h\to\tau\bar\mu$) decreases when increasing the heavy sparticle masses. 
This plot is for the particular input $Y_{\tau\mu}^{(1)}$, but similar behaviors (not shown) are obtained for the other two studied textures $Y_{\tau\mu}^{(2)}$ and  $Y_{\tau\mu}^{(3)}$.
 In this figure we have included the full predictions for BR($h\to\tau\bar\mu$), as well as the separated contributions coming only from chargino-sneutrino loops, i.e, diagrams (1)-(4) in fig.\ref{SUSY-diag}, and from neutralino-slepton loops, i.e., diagrams (5)-(8) in fig.\ref{SUSY-diag}. We see that not only the full prediction but also the separated contributions from these two subsets decrease with $m_{\rm SUSY}$, 
showing that the decoupling occurs in both, the charginos-sneutrinos and the neutralinos-sleptons sectors, as expected from the decoupling theorem. 
We also see that, in this heavy sparticles scenario, the contributions from the charginos-sneutrinos sector dominate by many orders of magnitude over the ones from the neutralinos-sleptons sector. 
In order to better understand the contributions from the charginos-sneutrinos sector, which are the ones containing the new sparticles with respect to the MSSM, we consider next the simple case of $\Delta m_{\tilde L_{ij}}=0$, where the contributions from the neutralinos-sleptons sector vanish, and only the contributions from charginos-sneutrinos remain. We show in fig.\ref{decoulingplots} right the separated contributions from each diagram (1), (2), (3), and (4) of fig.\ref{SUSY-diag}, and the full result. The contribution from diagram (1) is clearly subleading by several orders of magnitude and the contributions from the vertex correction, diagram (2), and the self-energies, diagrams (3) and (4), clearly compete in size. We also see that their  interference is destructive, such that the full result, that decouples with $m_{\rm SUSY}$, manifests that a strong cancellation among self-energies and vertex corrections is happening, as expected. Notice also that diagrams (3) and (4) do not decouple individually with $m_{\rm SUSY}$, but they do decouple when adding all the diagrams, as expected.

\begin{figure*}[t!]
\begin{center}
\begin{tabular}{cc}
\includegraphics[width=0.475\textwidth]{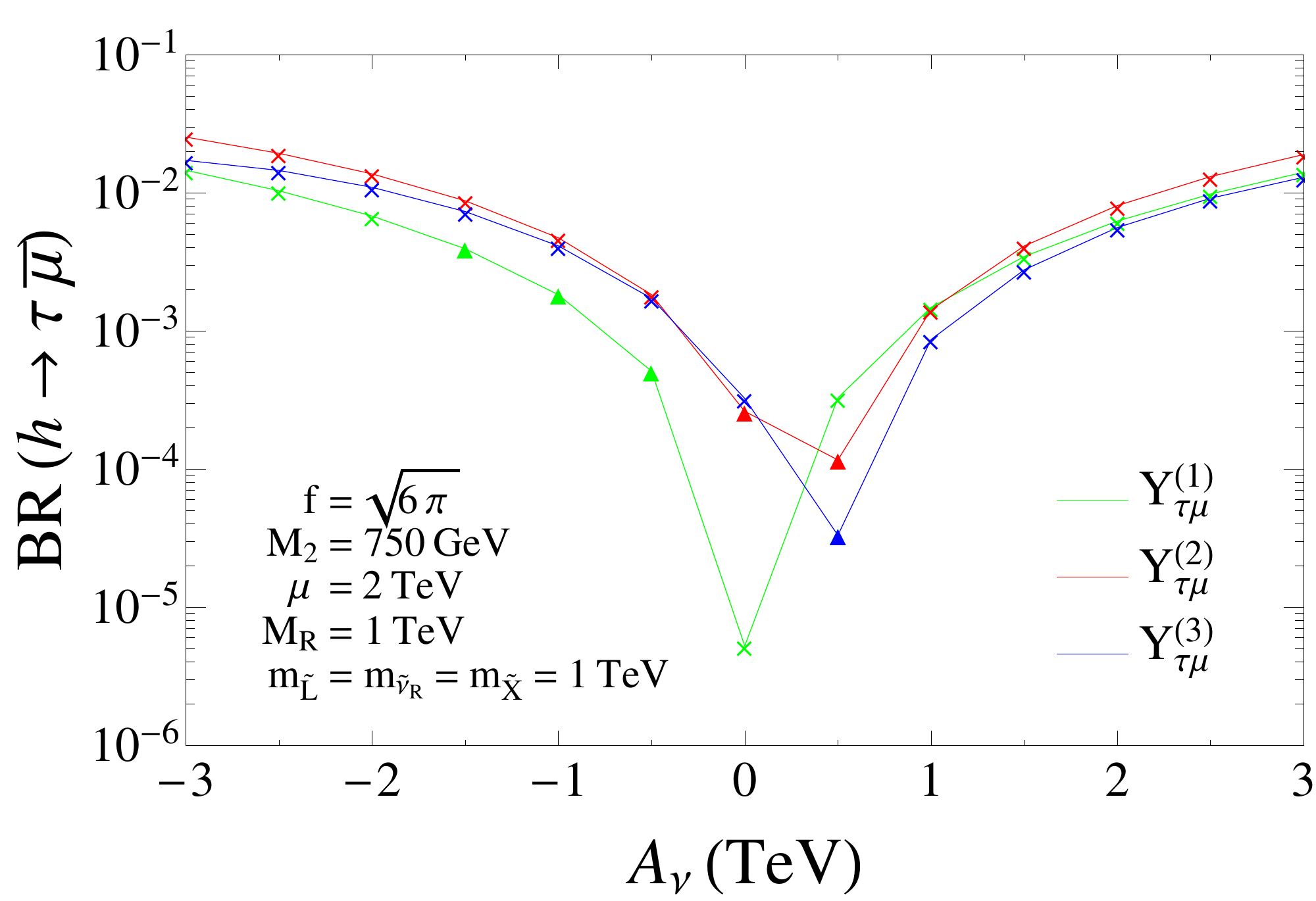}&
\includegraphics[width=0.475\textwidth]{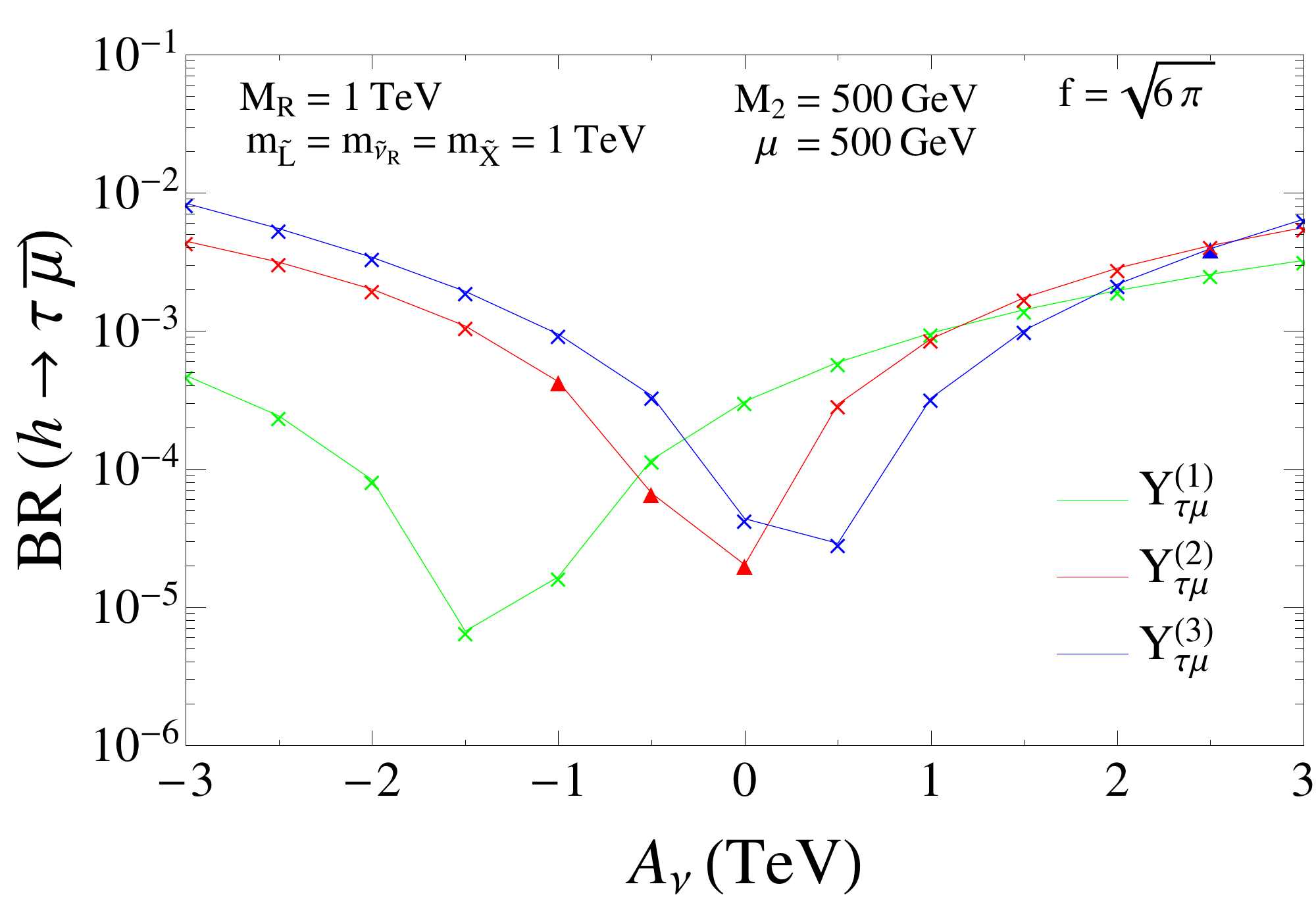}
\end{tabular}
\caption{Dependence of BR($h \to \tau \bar \mu$) on $A_\nu$ for the three neutrino Yukawa couplings $Y_{\tau\mu}^{(1)}$, $Y_{\tau\mu}^{(2)}$,
and $Y_{\tau\mu}^{(3)}$, with $M_2 =$ 750 GeV, $\tan\beta =$ 5 and $\mu =$ 2 TeV (left panel) or with $\tan\beta =$ 10 and $M_2 = \mu =$ 500 GeV (right panel).
On both panels, $m_A =$ 800 GeV, $M_0=$ 1 TeV, $M_R =$ $m_{\tilde L} =$ $m_{\tilde e} =$ $m_{\tilde \nu_R} =$ $m_{\tilde X} =$ 1 TeV, and the scaling factor $f = \sqrt{6\pi}$. Crosses (triangles) represent points
in the SUSY-ISS parameter space excluded (allowed) by the $\tau \to \mu \gamma$ upper limit, BR($\tau \to \mu \gamma$) $< 4.4\times 10^{-8}$~\cite{Aubert:2009ag}.}
\label{LFVHD-Anu}
\end{center}
\end{figure*}

Regarding the relevance for the searched enhancement in the SUSY contributions with respect to the trilinear coupling $A_\nu$, we have found that the LFVHD rates are indeed very
sensitive to the particular value of $A_\nu$. Thus, we study in figure~\ref{LFVHD-Anu}
the behavior
of BR($h \to \tau \bar \mu$) with this parameter for the two scenarios considered previously, with $M_R =$ 1 TeV and the scaling factor $f = \sqrt{6\pi}$. On both plots we confirm
the strong dependence of the LFVHD branching ratios with $A_\nu$, presenting deep dips in different positions that depend mainly on the values
of $Y_\nu$, $\mu$, $m_A$ and $\tan\beta$. In particular,
the $h^0-\tilde \nu_L-\tilde\nu_R$ coupling and the $\tilde\nu_L-\tilde\nu_R$ mixing are controlled by these parameters, which would lead to the appearance of dips in the regime where
contributions from sneutrino-chargino loops dominate. This is the case of figure~\ref{LFVHD-Anu} and it is interesting to note that, for this choice of parameters,
practically all the parameter space is excluded
by $\tau \to \mu \gamma$ except the points within the dips and surrounding them, where the LFV radiative decay $\tau \to \mu \gamma$ suffers also a strong reduction.
An interesting feature we found is that the location
of the dips in BR($h \to \tau \bar \mu$) and BR($\tau \to \mu \gamma$) usually do not coincide, therefore allowing for large LFV Higgs decays rates, not excluded by $\tau \to \mu \gamma$, above $10^{-3}$ and within the reach of
the LHC experiments.

\begin{figure*}[t!]
\begin{center}
\begin{tabular}{cc}
\includegraphics[width=0.475\textwidth]{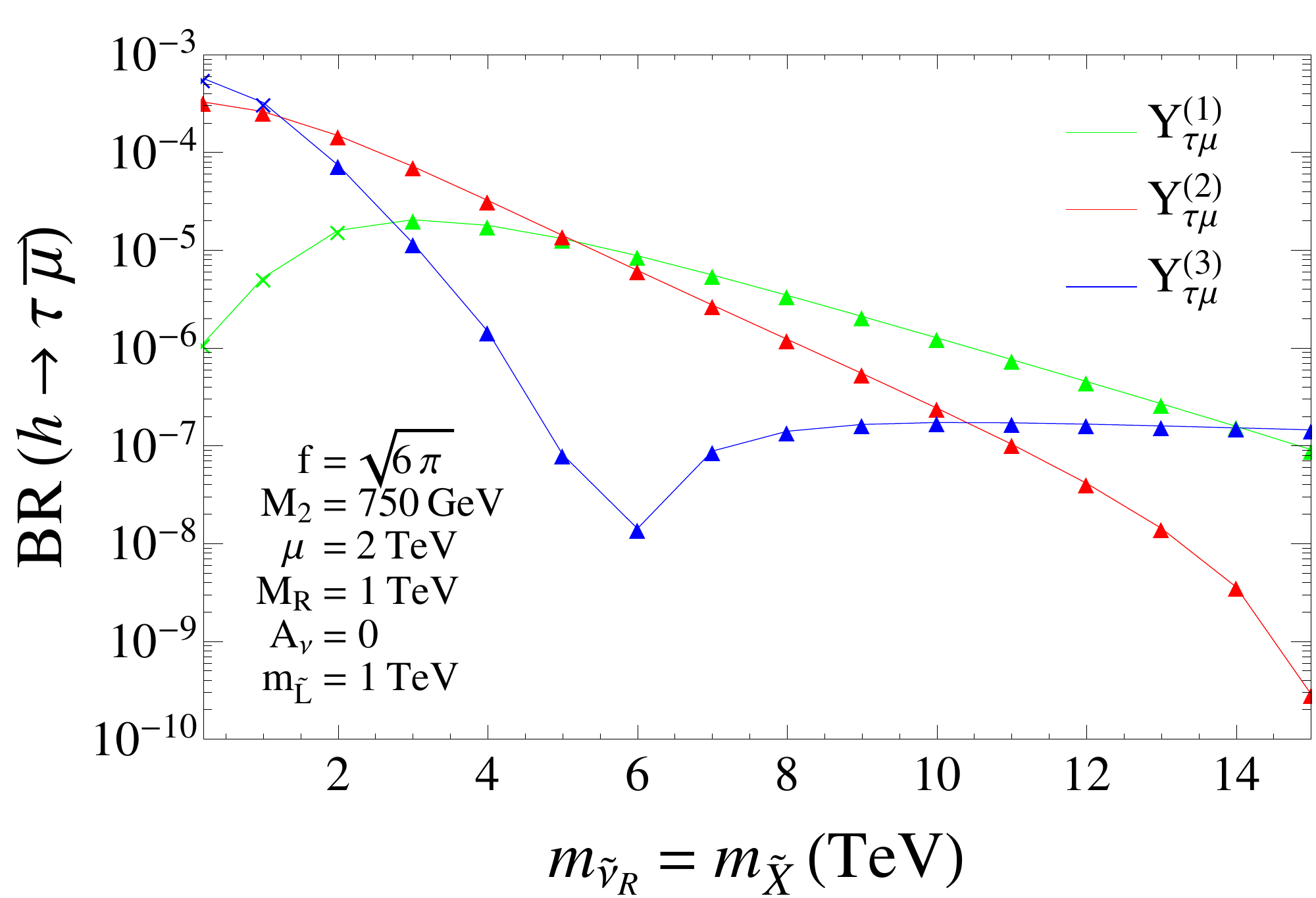}&
\includegraphics[width=0.475\textwidth]{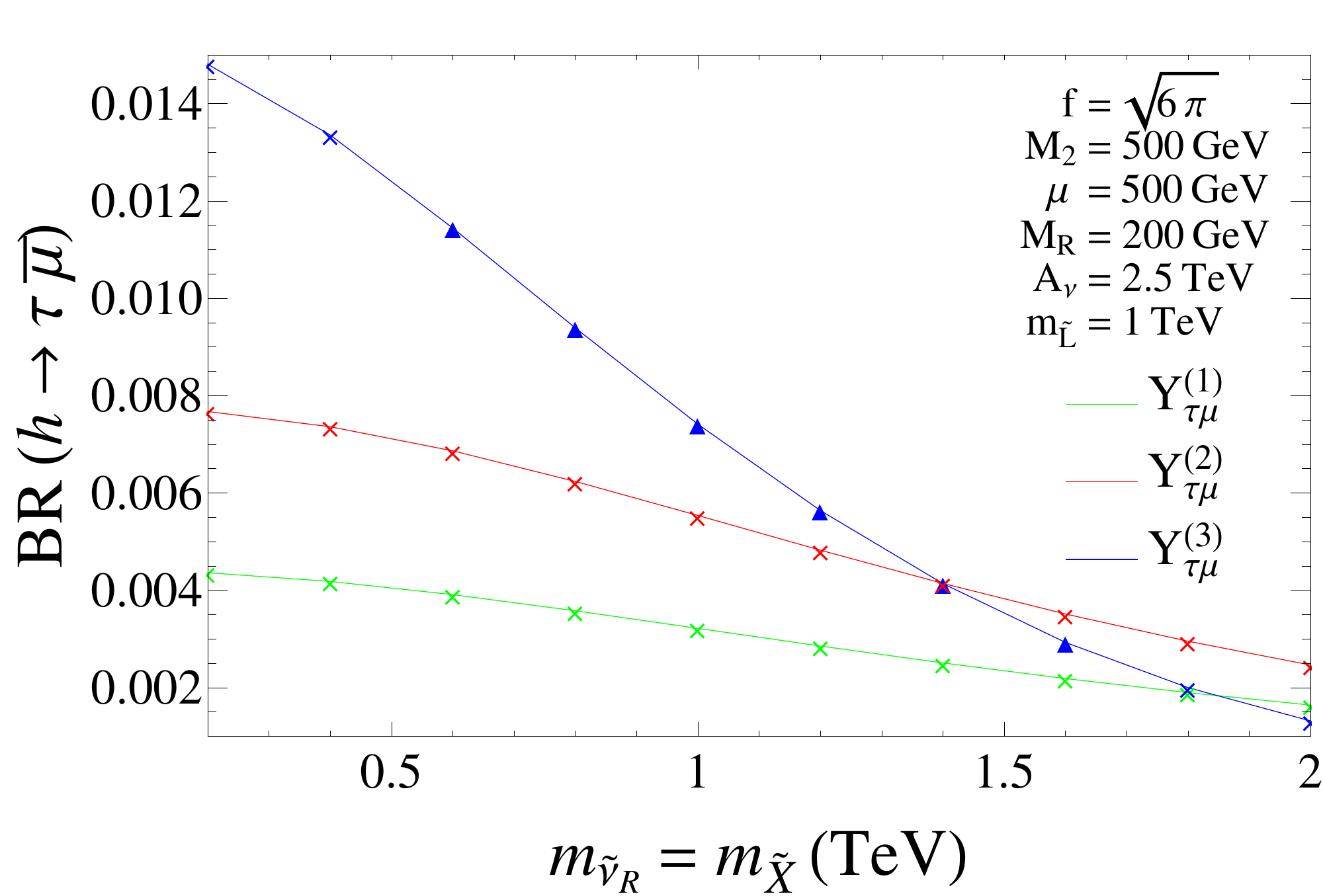}
\end{tabular}
\caption{Dependence of BR($h \to \tau \bar \mu$) on $m_{\tilde \nu_R}=m_{\tilde X}$ for the three neutrino Yukawa couplings $Y_{\tau\mu}^{(1)}$, $Y_{\tau\mu}^{(2)}$,
and $Y_{\tau\mu}^{(3)}$, with $M_R =$ $m_{\tilde L} =$ $m_{\tilde e} =$ 1 TeV, $M_2 =$ 750 GeV, $\mu =$ 2 TeV, $\tan\beta =$ 5 and $A_\nu =$ 0 (left panel) or
with $M_R =$ 200 GeV, $m_{\tilde L} =$ $m_{\tilde e} =$ 1 TeV, $M_2 = \mu =$ 500 GeV, $\tan\beta =$ 10 and $A_\nu =$ 2.5 TeV (right panel). On both
panels, $m_A =$ 800 GeV, $M_0=$ 1 TeV, and $f = \sqrt{6\pi}$. Crosses (triangles) represent points in the SUSY-ISS parameter space excluded (allowed)
by the $\tau \to \mu \gamma$ upper limit, BR($\tau \to \mu \gamma$) $< 4.4\times 10^{-8}$~\cite{Aubert:2009ag}.}
\label{LFVHD-msoft}
\end{center}
\end{figure*}

Finally, the dependence of the LFVHD rates on the new sneutrino soft SUSY breaking scalar masses, $m_{\tilde \nu_R}$ and $m_{\tilde X}$, is depicted in figure~\ref{LFVHD-msoft}
where these parameters are varied independently from the SUSY scale. As when varying $M_R$, increasing $m_{\tilde \nu_R}$ and $m_{\tilde X}$ makes the singlet sneutrinos heavier
and decreases
the size of the chargino contribution. For $Y_{\tau\mu}^{(1)}$ and $Y_{\tau\mu}^{(2)}$ which are dominated by this contribution, the BR($h \to \tau \bar \mu$) exhibits a
strong decrease between 200 GeV and 14 TeV, by more than five orders of magnitude in the case of $Y_{\tau\mu}^{(2)}$. For $Y_{\tau\mu}^{(3)}$ a dip can be observed, due again to a cancellation
between the chargino and neutralino contributions, with the latter dominating at large $m_{\tilde \nu_R}$. For the first benchmark point, the largest $h \to \tau \bar \mu$ rates allowed
by the $\tau \to \mu \gamma$ upper limit are obtained for $Y_{\tau\mu}^{(2)}$
with $m_{\tilde \nu_R} =$ 200 GeV, with a maximum value of $\sim 3 \times 10^{-4}$, just one order of magnitude below the present LHC sensitivity.
If we move our attention to the vicinity of the region of
low values of $m_{\tilde \nu_R}$ for the second benchmark point, we found large LFVHD rates, as displayed on the right panel of figure~\ref{LFVHD-msoft},
with $M_R =$ 200 GeV and $A_\nu =$ 2.5 TeV. We observe a huge increase in BR($h \to \tau \bar \mu$)
for the three Yukawa textures $Y_{\tau\mu}^{(1)}$, $Y_{\tau\mu}^{(2)}$, and $Y_{\tau\mu}^{(3)}$, with maximum values of $\sim 4 \times 10^{-3}$, $\sim 8 \times 10^{-3}$, and
$\sim 1.5 \times 10^{-2}$, respectively, due mainly to the low values of $m_{\tilde \nu_R}$ and $M_R$. Unfortunately, all the parameter space for $Y_{\tau\mu}^{(1)}$ and
$Y_{\tau\mu}^{(2)}$ cases is excluded by the $\tau \to \mu \gamma$ upper limit. By contrast, most of the points for the $Y_{\tau\mu}^{(3)}$ texture are in agreement with this upper
bound, because they are located in a region where the $\tau \to \mu \gamma$ rates suffer a strong suppression as a consequence of the value set for $A_\nu$ in
this case, $A_\nu =$ 2.5 TeV. This fact allows us to obtain a maximum value of BR($h \to \tau \bar \mu$) $\sim 1.1\%$, completely within 
the reach of the current LHC experiments and large enough to explain the CMS and ATLAS excesses if confirmed by other experiments and/or future data. 

\section{Conclusions}
\label{conclusions}
In this article, we have presented the results of an updated and full one-loop calculation of the SUSY contributions to lepton flavor violating Higgs decays in the SUSY-ISS
model. We found much larger contributions than in the type I seesaw
due to the lower values of $M_R \sim {\cal O}(1$ TeV), an increased RGE-induced slepton mixing, and the presence of right-handed
sneutrinos at the TeV scale, where both sleptons and sneutrinos large
couplings transmit sizable LFV due to the large $Y_\nu^2/(4 \pi) \sim  {\cal O} (1) $
considered here. We showed that the branching ratio of $h \to \tau \bar\mu$ exhibits different behaviors as a function of the seesaw and SUSY scale if it is dominated by
chargino or neutralino loops. Moreover, a nonzero trilinear coupling $A_\nu$ leads to increased LFVHD rates. Choosing different benchmark points, we found that 
BR($h \to \tau \mu$) of the order of $10^{-2}$ can be reached while agreeing with the experimental limits on radiative decays, providing a possible explanation of the CMS and ATLAS excesses. While out of the scope of this work, a complete study including nonsupersymmetric contributions in the SUSY-ISS model and a detailed analysis of experimental
constraints beyond radiative LFV decays will be presented in a future article.

\section*{Acknowledgments}

This work is supported by the European Union Grant No. FP7 ITN
INVISIBLES (Marie Curie Actions, Grant No. PITN-GA-2011-289442), by the CICYT through Grant No. FPA2012-31880,  
by the Spanish Consolider-Ingenio 2010 Programme CPAN (Grant No. CSD2007-00042), 
and by the Spanish MINECO's ``Centro de Excelencia Severo Ochoa'' Programme under Grant No. SEV-2012-0249.
E.~A. is financially supported by the Spanish DGIID-DGA Grant No. 2013-E24/2 and the Spanish MICINN Grants No. FPA2012-35453 and No. CPAN-CSD2007-00042.
X.~M. is supported through the FPU Grant No. AP-2012-6708. C.~W. received financial support as an International Research Fellow of the Japan Society for the Promotion of Science and
from the European Research Council under the European Union's Seventh Framework Programme (FP/2007-2013) / ERC Grant NuMass Agreement No. [617143] during different stages of this work.  

\end{multicols}

\section*{Appendices}
\appendix

\section{Mass matrices and couplings in the SUSY-ISS model \label{App_Couplings}}

We present in this appendix the mass matrices and coupling factors that are relevant to our calculation of the LFV Higgs decays. The sneutrino mass matrix $M^2_{\tilde \nu}$ is defined by
\begin{equation}\label{Msnu}
 -\mathcal{L}^{\tilde \nu}_{\mathrm{mass}}=\frac{1}{2}\left(\tilde \nu_L^\dagger\,, \tilde \nu_L^T\,, \tilde \nu_R^T\,, \tilde \nu_R^\dagger\,, \tilde X^T\,, \tilde X^\dagger\right) 
 M^2_{\tilde \nu} \left( \begin{array}{c} \tilde \nu_L \\ \tilde \nu_L^*\\ \tilde \nu_R^*\\ \tilde \nu_R \\ \tilde X^* \\ \tilde X \end{array} \right)\,,
\end{equation}
where $\tilde \nu_L$, $\tilde\nu_R$ and $\tilde X$ are vectors made of weak eigenstates and defined in a similar fashion,
e.g. $\tilde \nu_L= (\tilde \nu_L^{(e)}\,,\tilde \nu_L^{(\mu)}\,, \tilde \nu_L^{(\tau)})^T$. The $18\times18$ sneutrino mass matrix is expressed in terms of $3\times3$ submatrices, giving
\begin{equation}
 M^2_{\tilde \nu}=\left( \begin{array}{cccccc}
                          M^2_{LL} & 0 & 0 & M^2_{LR} & m_D M_R^* & 0 \\
                          0 & (M^2_{LL})^T & (M^2_{LR})^* & 0 & 0 & m_D^* M_R \\
                          0 & (M^2_{LR})^T & M^2_{RR} & 0 & M_R \mu_X^* & (B_R M_R^*)^* \\
                          (M^2_{LR})^\dagger & 0 & 0 & (M^2_{RR})^T & B_R M_R^* & M_R^* \mu_X \\
                          M_R ^T m_D^\dagger & 0 & \mu_X M_R^\dagger & (B_R M_R^*)^\dagger & M^2_{XX} & 2 (B_X \mu_X^*)^\dagger \\
                          0 & M_R^\dagger m_D^T & (B_R M_R^*)^T & \mu_X^* M_R^T & 2 (B_X \mu_X^*) & (M^2_{XX})^T
                         \end{array} \right)\,,
\end{equation}
with
\begin{align}
 M^2_{LL} &= m_D m_D^\dagger + m^2_{\tilde L} + \mathbb{1} \frac{m_Z}{2} \cos 2\beta\,, \\
 M^2_{LR} &= -\frac{\mu}{\tan \beta} m_D + m_D A_\nu^\dagger\,, \\
 M^2_{RR} &= m_D^T m_D^* + M_R M_R^\dagger + m^2_{\tilde \nu_R}\,, \\
 M^2_{XX} &= M_R^T M_R^* + \mu_X \mu_X^* + m^2_{\tilde X}\,,
\end{align}
where we have used the fact that $\mu_X$ is a symmetric matrix. Then, the sneutrino mass matrix is diagonalized using
\begin{equation}
 \tilde U^\dagger M^2_{\tilde \nu} \tilde U = M^2_{\tilde n} = \mathrm{diag}(m^2_{\tilde n_1}\,, ...\,, m^2_{\tilde n_{18}} ) \,,
\end{equation}
which corresponds to
\begin{equation}
 \left( \begin{array}{c} \tilde \nu_L \\ \tilde \nu_L^*\\ \tilde \nu_R^*\\ \tilde \nu_R \\ \tilde X^* \\ \tilde X \end{array} \right) =
 \tilde U \left( \begin{array}{c} \tilde n_1 \\ \vdots \\ \vdots \\ \vdots \\ \vdots \\ \tilde n_{18} \end{array} \right)\,. \label{snuRot}
\end{equation}
The basis in eq.~(\ref{Msnu}) uses the sneutrino electroweak eigenstates and their complex conjugate states, and they fulfill
\begin{align}
 \tilde \nu_i &= \tilde U_{i,j} \tilde n_j\,, \\
 \tilde \nu_i^* &= \tilde U_{3+i,j} \tilde n_j\,, \label{thisone}
\end{align}
and
\begin{equation}
 (\tilde \nu_i)^* = \tilde U_{i,j}^* \tilde n_j\,, \label{thistwo}
\end{equation}
since the physical sneutrinos are real scalar fields. While both eqs.~(\ref{thisone}) and (\ref{thistwo}) are equally valid, we choose eq.~(\ref{thisone}). The mass
matrices of the other SUSY particles, namely the charginos, neutralinos, and charged sleptons, are the same as in the SUSY type I seesaw studied in~\cite{Arganda:2004bz} and we will
use their definitions of the corresponding rotation matrices, which in turns were based on the conventions of~\cite{Gunion:1984yn} for the charginos and neutralinos. 
Concretely, $U$ and $V$ will be the matrices that rotate the chargino states and $N$ the one that rotates the neutralino states.
In addition, combinations of 
rotation matrices for the neutralinos are defined as
\begin{align}
 N'_{a1}&=N_{a1}\cos\theta_W+N_{a2}\sin\theta_W\,, \nonumber \\
 N'_{a2}&=-N_{a1}\sin\theta_W+N_{a2}\cos\theta_W\,. \nonumber \\
\end{align}
As for the charged sleptons, they are diagonalized by
\begin{equation}
 \tilde \ell^{'} = R^{(\ell)} \tilde \ell\,,
\end{equation}
where $\tilde \ell^{'}=(\tilde e_L,\tilde e_R,\tilde\mu_L,\tilde\mu_R,\tilde\tau_L\,, \tilde \tau_R)^T$ are the weak eigenstates
and $\tilde \ell=(\tilde \ell_1\,, ...\,, \tilde \ell_6)^T$ are the mass eigenstates.

When compared with the SUSY type I seesaw, only the coupling factors $A_{R\alpha j}^{(\ell)}$ and $g_{H_x \tilde{\nu}_{\alpha} \tilde{\nu}_{\beta}}$ are modified. In the SUSY inverse seesaw, they are defined in the mass basis with diagonal charged leptons by
\begin{align}
 A_{R\alpha j}^{(e,\mu,\tau)}=& \tilde U_{(1,2,3)\alpha} V_{j1} - \frac{{m_D}_{(1,2,3)k}}{\sqrt{2}m_W\sin \beta} \tilde U_{k+9,\alpha} V_{j2} \,, \nonumber \\
 g_{H_x \tilde{\nu}_{\alpha} \tilde{\nu}_{\beta}} = &-\imath g \left[ (g_{LL, \nu}^{(x)})_{ik} \tilde U_{i\alpha}^{*} \tilde U_{k\beta} 
    + (g_{RR, \nu}^{(x)})_{ik} \tilde U_{i+9,\alpha}^{*} \tilde U_{k+9,\beta} \right. \nonumber \\
  &+ \left. (g_{LR, \nu}^{(x)})_{ik} \tilde U_{i,\alpha}^{*} \tilde U_{k+9,\beta} + (g_{LR, \nu}^{(x)})^*_{ik} \tilde U^*_{k+9,\alpha} \tilde U_{i,\beta}\right. \nonumber \\
  &+ \left. (g_{LX, \nu}^{(x)})_{ik} \tilde U_{i,\alpha}^{*} \tilde U_{k+12,\beta} + (g_{LX, \nu}^{(x)})^*_{ik} \tilde U^*_{k+12,\alpha} \tilde U_{i,\beta} \right]\,, \nonumber\\
 (g_{LL, \nu}^{(x)})_{ik} =&- \frac{m_Z}{2\cos{\theta_W}} \sigma_3^{(x)} \delta_{ik} +\frac{(m_D m_D^\dagger)_{ik}}{m_W \sin \beta} \sigma_6^{(x)}\,, \nonumber \\
 (g_{RR, \nu}^{(x)})_{ik} =& \frac{(m_D^\dagger m_D)_{ik}}{m_W \sin\beta} \sigma_6^{(x)}\,, \nonumber \\
 (g_{LR, \nu}^{(x)})_{ik} =&   \frac{(m_D A_\nu^\dagger)_{ik}}{2m_W\sin\beta} \sigma_2^{(x)} + \frac{\mu}{ 2m_W\sin \beta} (m_D)_{ik} \sigma_7^{(x)} \,, \nonumber \\
 (g_{LX, \nu}^{(x)})_{ik} =&  \frac{(m_D M_R^*)_{ik}}{2 m_W \sin\beta} \sigma_2^{(x)}\,,
\end{align}
which are summed over the internal indices, with $i\,,k=1\,,...\,,3$. We reproduced below the unmodified coupling factors from~\cite{Arganda:2004bz} (correcting a typo in $W_{Rij}^{(x)}$) 
for completeness in the mass basis with diagonal charged leptons
\begin{align}
A_{L\alpha j}^{(e,\mu,\tau)}=& -\frac{m_{e,\mu,\tau}}{\sqrt{2}m_W cos\beta}U_{j2}^* \tilde U_{(1,2,3)\alpha}\,, \nonumber \\
B_{L\alpha a}^{(e,\mu,\tau)}=& \sqrt{2}\left[\frac{m_{e,\mu,\tau}}{2m_W cos\beta}N_{a3}^*R_{(1,3,5)\alpha}^{(\ell)} +\left[\sin\theta_W 
  N_{a1}^{'*}-\frac{\sin^2\theta_W}{cos\theta_W}N_{a2}^{'*}\right]R_{(2,4,6)\alpha}^{(\ell)}\right]\,, \nonumber \\
B_{R\alpha a}^{(e,\mu,\tau)}=& \sqrt{2}\left[\left(-\sin\theta_WN_{a1}^{'}-\frac{1}{\cos\theta_W}(\frac{1}{2}-\sin^2\theta_W)N_{a2}^{'}\right)
  R_{(1,3,5)\alpha}^{(\ell)}+ \frac{m_{e,\mu,\tau}}{2m_W \cos\beta}N_{a3}R_{(2,4,6)\alpha}^{(\ell)}\right]\,, \nonumber \\
W_{Lij}^{(x)}=&\frac{1}{\sqrt{2}}\left(-\sigma_1^{(x)}U_{j2}^*V_{i1}^*+ \sigma_2^{(x)}U_{j1}^*V_{i2}^*\right)\,, \nonumber \\
W_{Rij}^{(x)}=&\frac{1}{\sqrt{2}}\left(-\sigma_1^{(x)*}U_{i2}V_{j1}+\sigma_2^{(x)*}U_{i1}V_{j2}\right)\,, \nonumber \\
D_{Lab}^{(x)}=&\frac{1}{2\cos\theta_W}\left[(\sin\theta_W N_{b1}^*-\cos\theta_W N_{b2}^*)(\sigma_1^{(x)}N_{a3}^*+\sigma_2^{(x)}N_{a4}^*) \right. \nonumber \\
  &+(\sin\theta_W N_{a1}^*-\cos\theta_W   N_{a2}^*)(\sigma_1^{(x)}N_{b3}^*+\sigma_2^{(x)}N_{b4}^*)\left.\right] \,, \nonumber \\
D_{Rab}^{(x)}=&D_{Lab}^{(x)*}\,, \nonumber \\
S_{L,\ell}^{(x)} =& - \frac{m_{\ell}}{2 m_W \cos\beta}{\sigma_1^{(x)*}}\,, \nonumber \\
S_{R,\ell}^{(x)} =&S_{L, \ell}^{(x)*} \,, \nonumber \\
g_{H_x \tilde{\ell}_{\alpha} \tilde{\ell}_{\beta}} =& -\imath g \left[ g_{LL, e}^{(x)} R_{1\alpha}^{*(\ell)} R_{1\beta}^{(\ell)} + 
   g_{RR, e}^{(x)} R_{2\alpha}^{*(\ell)} R_{2\beta}^{(\ell)} + g_{LR, e}^{(x)} R_{1\alpha}^{*(\ell)} R_{2\beta}^{(\ell)} + g_{RL, e}^{(x)} R_{2\alpha}^{*(\ell)} R_{1\beta}^{(\ell)}\right. \nonumber \\
  &+ g_{LL, \mu}^{(x)} R_{3\alpha}^{*(\ell)} R_{3\beta}^{(\ell)} + g_{RR, \mu}^{(x)} R_{4\alpha}^{*(\ell)} R_{4\beta}^{(\ell)} + g_{LR, \mu}^{(x)} R_{3\alpha}^{*(\ell)} R_{4\beta}^{(\ell)} 
   + g_{RL, \mu}^{(x)} R_{4\alpha}^{*(\ell)} R_{3\beta}^{(\ell)} \nonumber \\
  &+ \left. g_{LL, \tau}^{(x)} R_{5\alpha}^{*(\ell)} R_{5\beta}^{(\ell)} + g_{RR, \tau}^{(x)} R_{6\alpha}^{*(\ell)} R_{6\beta}^{(\ell)} + 
   g_{LR, \tau}^{(x)} R_{5\alpha}^{*(\ell)} R_{6\beta}^{(\ell)} + g_{RL, \tau}^{(x)} R_{6\alpha}^{*(\ell)} R_{5\beta}^{(\ell)} \right] \,, \nonumber \\
g_{LL, \ell}^{(x)} =&  \frac{m_Z}{\cos{\theta_W}} \sigma_3^{(x)} \left( \frac{1}{2}- \sin^2{\theta_W} \right) + \frac{m_{\ell}^2}{m_W \cos{\beta}} \sigma_4^{(x)}\,, \nonumber\\
g_{RR, \ell}^{(x)} =&  \frac{m_Z}{\cos{\theta_W}} \sigma_3^{(x)} \left(  \sin^2{\theta_W} \right) + \frac{m_{\ell}^2}{m_W \cos{\beta}}  \sigma_4^{(x)}\,,\nonumber \\
g_{LR, \ell}^{(x)} =& \left(-\sigma_1^{(x)}A_\ell-\sigma_5^{(x)}\mu\right) \frac{m_{\ell}}{2 m_W \cos{\beta}}\,, \nonumber \\
g_{RL, \ell}^{(x)} =& g_{LR, \ell}^{(x)*}\,,
\end{align}
with
\begin{align}
&\sigma_1^{(x)} = \begin{pmatrix}
  \sin  \alpha   \\
  -\cos \alpha   \\
  \imath \sin \beta 
\end{pmatrix}, \quad
\sigma_2^{(x)} = \begin{pmatrix}
 \cos \alpha  \\
 \sin \alpha  \\
 -\imath \cos \beta
\end{pmatrix}, \quad
\sigma_3^{(x)} = \begin{pmatrix}
 \sin  ( \alpha  +  \beta) \\
 -\cos ( \alpha  +  \beta) \\
 0
\end{pmatrix}, \quad
\sigma_4^{(x)} = \begin{pmatrix}
 -\sin \alpha  \\
 \cos  \alpha  \\
 0
\end{pmatrix}, \quad
\nonumber \\
&\sigma_5^{(x)} = \begin{pmatrix}
 \cos \alpha  \\
 \sin \alpha  \\
 \imath \cos\beta
\end{pmatrix}, \quad
\sigma_6^{(x)} = \begin{pmatrix}
 \cos \alpha  \\
 \sin \alpha  \\
 0
\end{pmatrix}, \quad
\sigma_7^{(x)} = \begin{pmatrix}
 \sin    \alpha \\
 -\cos   \alpha \\
 -\imath \sin  \beta 
\end{pmatrix}, \quad
\textrm{ for } 
 H_x=\begin{pmatrix}
 h^0\\
 H^0\\
 A^0
\end{pmatrix}. 
\end{align}

\section{Form factors in the SUSY-ISS model \label{App_FF}}

We present here the form factors that correspond to the diagrams of figure~\ref{SUSY-diag}. The original calculation in the SUSY type I seesaw was carried by some of the
authors in the mass
basis and in the Feynman--'t Hooft gauge~\cite{Arganda:2004bz}. The only changes required to adapt the original form factors to the SUSY-ISS model are the sum over sneutrinos that has to be extended to the
18 mass eigenstates and the new couplings defined in Appendix~\ref{App_Couplings}. In the following formulas, summation over all indices corresponding to internal propagators is understood. These would be $\alpha\,, \beta=1\,,...\,,18$
for the sneutrinos, $i\,, j=1\,,2$ for the charginos, $\alpha\,, \beta=1\,,...\,,6$ for the charged sleptons and $a\,,b=1\,,...\,,4$ for the neutralinos.
\begin{eqnarray*}
F_{L,x}^{(1)} &=& - \frac{g^2}{16 \pi^2}\left[\left(B_0 + 
m_{\tilde {\nu}_{\alpha}}^2 C_0+m_{\ell_m}^2 C_{12}+m_{\ell_k}^2 (C_{11} - C_{12})\right)\,
\kappa_{L 1}^{x,\,\tilde \chi^-} \right.\nonumber \\
&&+m_{\ell_k} m_{\ell_m} \left(C_{11}+C_0\right)\kappa_{L 2}^{x, \tilde \chi^-}+
m_{\ell_k} m_{\tilde \chi _j^-} \left(C_{11}-C_{12}+C_0\right)\kappa_{L 3}^{x, \tilde \chi^-}\,+ 
m_{\ell_m} m_{\tilde \chi _j^-} C_{12}\,\kappa_{L 4}^{x,\tilde \chi^-}\nonumber \\
&& +m_{\ell_k} m_{\tilde \chi _i^-} \left(C_{11}-C_{12}\right)\kappa_{L 5}^{x,\tilde \chi^-} + 
m_{\ell_m} m_{\tilde \chi _i^-} \left(C_{12}+C_0\right)\kappa_{L 6}^{x,\tilde \chi^-}+ 
m_{\tilde \chi _i^-} m_{\tilde \chi _j^-}C_0\,\kappa_{L 7}^{x,\tilde \chi^-} \Big]\,, \nonumber \\
F_{L,x}^{(2)} &=&
- \frac{igg_{H_x\tilde {\nu}_\alpha \tilde {\nu}_\beta}}{16\pi^2}
\left[-m_{\ell_k}(C_{11}-C_{12})\,\iota_{L 1}^{x,\tilde \chi^-}-
m_{\ell_m} C_{12}\, \iota_{L 2}^{x,\tilde \chi^-}+
m_{\tilde \chi^-_i}C_0\,\iota_{L 3}^{x,\tilde \chi^-}\right]\,,\nonumber \\
F_{L,x}^{(3)} &=&
- \frac{S_{L,\ell_{m}}^{(x)}}{m_{\ell_k}^2-m_{\ell_m}^2}\left[m_{\ell_k}^2 \Sigma_R^{\tilde
    \chi^-} (m_{\ell_k}^2)+m_{\ell_k}^2 \Sigma_{Rs}^{\tilde \chi^-}(m_{\ell_k}^2)
+ m_{\ell_m}\left(m_{\ell_k} \Sigma_L^{\tilde \chi^-} (m_{\ell_k}^2)+
m_{\ell_k}\Sigma_{Ls}^{\tilde \chi^-} (m_{\ell_k}^2)\right)\right]\,,\nonumber \\
F_{L,x}^{(4)} &=&
- \frac{S_{L,\ell_{k}}^{(x)}}{m_{\ell_m}^2-m_{\ell_k}^2}
\left[m_{\ell_m}^2 \Sigma_L^{\tilde \chi^-} (m_{\ell_m}^2)+
m_{\ell_m} m_{\ell_k} \Sigma_{Rs}^{\tilde \chi^-}(m_{\ell_m}^2)\right.
+ \left. m_{\ell_k}\left(m_{\ell_m} \Sigma_R^{\tilde \chi^-} (m_{\ell_m}^2)
+m_{\ell_k} \Sigma_{Ls}^{\tilde \chi^-} (m_{\ell_m}^2)\right)\right]\,,\nonumber \\
F_{L,x}^{(5)} &=& - \frac{g^2}{16 \pi^2}\left[\left(B_0 + 
m_{\tilde \ell_{\alpha}}^2 C_0+m_{\ell_m}^2 C_{12}+m_{\ell_k}^2 (C_{11} - C_{12})\right)\,
\kappa_{L 1}^{x,\,\tilde \chi^0} \right.\nonumber \\
&&+m_{\ell_k} m_{\ell_m} \left(C_{11}+C_0\right)\kappa_{L 2}^{x, \tilde \chi^0}+
m_{\ell_k} m_{\tilde \chi _b^0} \left(C_{11}-C_{12}+C_0\right)\kappa_{L 3}^{x, \tilde \chi^0}\,+ 
m_{\ell_m} m_{\tilde \chi _b^0} C_{12}\,\kappa_{L 4}^{x,\tilde \chi^0}\nonumber \\
&& +m_{\ell_k} m_{\tilde \chi _a^0} \left(C_{11}-C_{12}\right)\kappa_{L 5}^{x,\tilde \chi^0} + 
m_{\ell_m} m_{\tilde \chi _a^0} \left(C_{12}+C_0\right)\kappa_{L 6}^{x,\tilde \chi^0}+ 
m_{\tilde \chi _a^0} m_{\tilde \chi _b^0}C_0\,\kappa_{L 7}^{x,\tilde \chi^0} \Big]\,, \nonumber \\
F_{L,x}^{(6)} &=&
- \frac{igg_{H_x\tilde \ell_\alpha \tilde \ell_\beta}}{16\pi^2}
\left[-m_{\ell_k}(C_{11}-C_{12})\,\iota_{L 1}^{x,\tilde \chi^0}-
m_{\ell_m} C_{12}\, \iota_{L 2}^{x,\tilde \chi^0}+
m_{\tilde \chi_a^0}C_0\,\iota_{L 3}^{x,\tilde \chi^0}\right]\,,\nonumber \\
F_{L,x}^{(7)} &=&
- \frac{S_{L,\ell_m}^{(x)}}{m_{\ell_k}^2-m_{\ell_m}^2}\left[m_{\ell_k}^2 \Sigma_R^{\tilde
    \chi^0} (m_{\ell_k}^2)+m_{\ell_k}^2 \Sigma_{Rs}^{\tilde \chi^0}(m_{\ell_k}^2)
+ m_{\ell_m}\left(m_{\ell_k} \Sigma_L^{\tilde \chi^0} (m_{\ell_k}^2)+
m_{\ell_k}\Sigma_{Ls}^{\tilde \chi^0} (m_{\ell_k}^2)\right)\right]\,,\nonumber \\
F_{L,x}^{(8)} &=&
- \frac{S_{L,\ell_k}^{(x)}}{m_{\ell_m}^2-m_{\ell_k}^2}
\left[m_{\ell_m}^2 \Sigma_L^{\tilde \chi^0} (m_{\ell_m}^2)+
m_{\ell_m} m_{\ell_k} \Sigma_{Rs}^{\tilde \chi^0}(m_{\ell_m}^2)\right.
+ \left. m_{\ell_k}\left(m_{\ell_m} \Sigma_R^{\tilde \chi^0} (m_{\ell_m}^2)
+m_{\ell_k} \Sigma_{Ls}^{\tilde \chi^0} (m_{\ell_m}^2)\right)\right]\,,\nonumber 
\label{formfactorLbs}
\end{eqnarray*}
where,
 \[B_0=
 \left\{   \begin{array}{l}
            B_0(m_{H_x}^2,m_{\tilde \chi^-_i}^2,m_{\tilde \chi^-_j}^2) \textrm{ in } F_{L,x}^{(1)}\,,\\
            B_0(m_{H_x}^2,m_{\tilde \chi^0_a}^2,m_{\tilde \chi^0_b}^2) \textrm{ in } F_{L,x}^{(5)}\,,
  \end{array}
\right. 
\]
and
 \[C_{0,11,12}=
 \left\{   \begin{array}{l}
C_{0,11,12} (m_{\ell_k}^2,m_{H_x}^2, m_{\tilde {\nu}_ {\alpha}}^2,m_{\tilde \chi^-_i}^2,m_{\tilde \chi^-_j}^2) \textrm{ in } F_{L,x}^{(1)}\,,\\
C_{0,11,12} (m_{\ell_k}^2,m_{H_x}^2, m_{\tilde \chi^-_i}^2,m_{\tilde {\nu}_ {\alpha}}^2,m_{\tilde {\nu}_ {\beta}}^2) \textrm{ in } F_{L,x}^{(2)}\,, \\
C_{0,11,12} (m_{\ell_k}^2,m_{H_x}^2, m_{\tilde {l}_ {\alpha}}^2,m_{\tilde \chi^0_a}^2,m_{\tilde \chi^0_b}^2) \textrm{ in } F_{L,x}^{(5)}\,,  \\
C_{0,11,12} (m_{\ell_k}^2,m_{H_x}^2, m_{\tilde \chi^0_a}^2,m_{\tilde {l}_ {\alpha}}^2,m_{\tilde {l}_ {\beta}}^2) \textrm{ in } F_{L,x}^{(6)}\,.
\end{array}
\right. 
\]

The couplings and self-energies from the neutralino contributions to the form factors were defined as
\begin{eqnarray*}
\kappa_{L 1}^{x,\,\tilde \chi^0} = B_{L \alpha a}^{(\ell_k)}D_{Rab}^{(x)}B_{R \alpha b}^{(\ell_m)*}\,, 
     && \iota_{L 1}^{x, \tilde\chi^0} = B_{R \alpha a}^{(\ell_k)}B_{R \beta a}^{(\ell_m)*}\,,\\
\kappa_{L 2}^{x,\,\tilde \chi^0} = B_{R \alpha a}^{(\ell_k)}D_{Lab}^{(x)}B_{L \alpha b}^{(\ell_m)*} \,,
     && \iota_{L 2}^{x, \tilde\chi^0} = B_{L \alpha a}^{(\ell_k)}B_{L \beta a}^{(\ell_m)*}\,,\\
\kappa_{L 3}^{x,\,\tilde \chi^0} = B_{R \alpha a}^{(\ell_k)}D_{Lab}^{(x)}B_{R \alpha b}^{(\ell_m)*}\,,
     && \iota_{L 3}^{x, \tilde\chi^0} = B_{L \alpha a}^{(\ell_k)}B_{R \beta a}^{(\ell_m)*}\,,\\
\kappa_{L 4}^{x,\,\tilde \chi^0} = B_{L \alpha a}^{(\ell_k)}D_{Rab}^{(x)}B_{L \alpha b}^{(\ell_m)*}\,, && \\
\kappa_{L 5}^{x,\,\tilde \chi^0} = B_{R \alpha a}^{(\ell_k)}D_{R ab}^{(x)}B_{R \alpha b}^{(\ell_m)*}\,, &&\\
\kappa_{L 6}^{x,\,\tilde \chi^0} = B_{L \alpha a}^{(\ell_k)}D_{L ab}^{(x)}B_{L \alpha b}^{(\ell_m)*}\,, &&\\
\kappa_{L 7}^{x,\,\tilde \chi^0} = B_{L \alpha a}^{(\ell_k)}D_{Lab}^{(x)}B_{R \alpha b}^{(\ell_m)*}\,, &&
\end{eqnarray*}\vspace*{-0.7cm}
\begin{eqnarray}
\Sigma_L^{\tilde\chi^0} (k^2) &=& -\frac{g^2}{16\pi^2}B_1(k^2,m_{\tilde\chi_a^0}^2, m_{\tilde \ell_\alpha}^2) B_{R\,\alpha a}^{(\ell_k)}B_{R\,\alpha a}^{(\ell_m)*}\,, \nonumber \\
m_{\ell_k} \Sigma_{Ls}^{\tilde \chi^0} (k^2) &=&  \frac{g^2m_{\tilde\chi_a^0}}{16\pi^2}B_0(k^2,m_{\tilde\chi_a^0}^2, m_{\tilde \ell_\alpha}^2) B_{L\,\alpha a}^{(\ell_k)}B_{R\,\alpha a}^{(\ell_m)*}\,.
\end{eqnarray}

The couplings and self-energies from the chargino contributions to the form factors,  $\kappa^{x, \, \tilde\chi^-}$, $\iota^{x, \, \tilde\chi^-}$, and $\Sigma^{\tilde\chi^-}$  can
be obtained from the previous expressions $\kappa^{x, \, \tilde\chi^0}$,  $\iota^{x, \, \tilde\chi^0}$ and $\Sigma^{ \tilde\chi^0}$ by using the following replacement rules
$m_{\tilde\chi_a^0}\rightarrow m_{\tilde\chi_i^-}$, $m_{\tilde \ell_\alpha}\rightarrow m_{\tilde \nu_\alpha}$, $B^{(l)}\rightarrow A^{(l)}$, 
$D^{(x)}\rightarrow W^{(x)}$, $a \rightarrow i$, and $b \rightarrow j$.
 
The form factors $F_{R,x}^{(i)},i=1,...,8$ can be obtained from $F_{L,x}^{(i)},i=1,...,8$ through the exchange $L\leftrightarrow R$ in all places.

\begin{multicols}{2}

\bibliographystyle{unsrt}

\begin{thebibliography}{99}

\bibitem{Aad:2012tfa}
  G.~Aad {\it et al.}  [ATLAS Collaboration],
  Phys.\ Lett.\ B {\bf 716} (2012) 1
  [arXiv:1207.7214 [hep-ex]].

\bibitem{Chatrchyan:2012ufa}
  S.~Chatrchyan {\it et al.}  [CMS Collaboration],
  Phys.\ Lett.\ B {\bf 716} (2012) 30
  [arXiv:1207.7235 [hep-ex]].

\bibitem{Aad:2015zhl}
  G.~Aad {\it et al.}  [ATLAS and CMS Collaborations],
  Phys.\ Rev.\ Lett.\  {\bf 114} (2015) 191803
  [arXiv:1503.07589 [hep-ex]].
  
\bibitem{Khachatryan:2015kon}
  V.~Khachatryan {\it et al.} [CMS Collaboration],
  Phys.\ Lett.\ B {\bf 749} (2015) 337
  [arXiv:1502.07400 [hep-ex]].

\bibitem{Aad:2015gha}
  G.~Aad {\it et al.} [ATLAS Collaboration],
  JHEP {\bf 1511} (2015) 211
  doi:10.1007/JHEP11(2015)211
  [arXiv:1508.03372 [hep-ex]].
  
\bibitem{Pilaftsis:1992st}
  A.~Pilaftsis,
  Phys.\ Lett.\ B {\bf 285} (1992) 68.
  
\bibitem{Arganda:2004bz}
  E.~Arganda, A.~M.~Curiel, M.~J.~Herrero and D.~Temes,
  Phys.\ Rev.\ D {\bf 71} (2005) 035011
  [hep-ph/0407302].

\bibitem{ISSrefs}  
  R.~N.~Mohapatra,
  Phys.\ Rev.\ Lett.\  {\bf 56} (1986) 561;
  R.~N.~Mohapatra and J.~W.~F.~Valle,
  Phys.\ Rev.\ D {\bf 34} (1986) 1642;
  J.~Bernabeu, A.~Santamaria, J.~Vidal, A.~Mendez and J.~W.~F.~Valle,
  Phys.\ Lett.\ B {\bf 187} (1987) 303.

\bibitem{Arganda:2014dta}
  E.~Arganda, M.~J.~Herrero, X.~Marcano and C.~Weiland,
  Phys.\ Rev.\ D {\bf 91} (2015) 015001
  [arXiv:1405.4300 [hep-ph]].
  
\bibitem{LFVHDsusy}
See, for example,
  J.~L.~Diaz-Cruz,
  JHEP {\bf 0305} (2003) 036
  [hep-ph/0207030];
  A.~Brignole and A.~Rossi,
  Phys.\ Lett.\ B {\bf 566} (2003) 217
  [hep-ph/0304081];
  A.~Brignole and A.~Rossi,
  Nucl.\ Phys.\ B {\bf 701} (2004) 3
  [hep-ph/0404211];
  A.~Vicente,
  Adv.\ High Energy Phys.\  {\bf 2015} (2015) 686572
  [arXiv:1503.08622 [hep-ph]].
  
\bibitem{Arana-Catania:2013xma}
  M.~Arana-Catania, E.~Arganda and M.~J.~Herrero,
  JHEP {\bf 1309} (2013) 160 [JHEP {\bf 1510} (2015) 192]
  [arXiv:1304.3371 [hep-ph]].

\bibitem{Abada:2011hm}
  A.~Abada, D.~Das and C.~Weiland,
  JHEP {\bf 1203} (2012) 100
  [arXiv:1111.5836 [hep-ph]].
  
\bibitem{Abada:2014kba}
  A.~Abada, M.~E.~Krauss, W.~Porod, F.~Staub, A.~Vicente and C.~Weiland,
  JHEP {\bf 1411} (2014) 048
  [arXiv:1408.0138 [hep-ph]].
  
\bibitem{Arganda:2015ija}
  E.~Arganda, M.~J.~Herrero, X.~Marcano and C.~Weiland,
  Phys.\ Lett.\ B {\bf 752} (2016) 46
  doi:10.1016/j.physletb.2015.11.013
  [arXiv:1508.05074 [hep-ph]].
  
\bibitem{Hisano:1995cp}
  J.~Hisano, T.~Moroi, K.~Tobe and M.~Yamaguchi,
  Phys.\ Rev.\ D {\bf 53} (1996) 2442
  [hep-ph/9510309].
  
\bibitem{GonzalezGarcia:1988rw}
  M.~C.~Gonzalez-Garcia and J.~W.~F.~Valle,
  Phys.\ Lett.\ B {\bf 216} (1989) 360.

\bibitem{PMNS}
  B.~Pontecorvo,
  Sov.\ Phys.\ JETP {\bf 6} (1957) 429
   [Zh.\ Eksp.\ Teor.\ Fiz.\  {\bf 33} (1957) 549].
  Z.~Maki, M.~Nakagawa and S.~Sakata,
  Prog.\ Theor.\ Phys.\  {\bf 28} (1962) 870.



\bibitem{Adam:2013mnn}
  J.~Adam {\it et al.} [MEG Collaboration],
  Phys.\ Rev.\ Lett.\  {\bf 110} (2013) 201801
  [arXiv:1303.0754 [hep-ex]].
  
\bibitem{Aubert:2009ag}
  B.~Aubert {\it et al.}  [BaBar Collaboration],
  Phys.\ Rev.\ Lett.\  {\bf 104} (2010) 021802
  [arXiv:0908.2381 [hep-ex]].
  
\bibitem{Gunion:1984yn}
  J.~F.~Gunion and H.~E.~Haber,
  Nucl.\ Phys.\ B {\bf 272} (1986) 1
   [Nucl.\ Phys.\ B {\bf 402} (1993) 567].
 
\end{thebibliography}

\end{multicols}

\end{document}